# Rosetta Statements: Simplifying FAIR Knowledge Graph Construction with a User-Centered Approach

Vogt, Lars[1] orcid.org/0000-0002-8280-0487;

Farfar, Kheir Eddine[1] orcid.org/0000-0002-0366-4596

Karanth, Pallavi[1] https://orcid.org/0009-0007-5934-1087

Konrad, Marcel[1] orcid.org/0000-0002-2452-3143

Oelen, Allard[1] orcid.org/0000-0001-9924-9153

Prinz, Manuel[1] orcid.org/0000-0003-2151-4556

Strömert, Philip[1] orcid.org/0000-0002-1595-3213

[1] *TIB Leibniz Information Centre for Science and Technology, Welfengarten 1B, 30167 Hanover, Germany*

Corresponding Author: lars.m.vogt@googlemail.com

# Abstract

Machines need machine-actionable, and thus FAIR (findable, accessible, interoperable, reusable) data and metadata to support researchers in managing exponentially increasing amounts of data. Knowledge graphs, ontologies, and semantic graph patterns are promising technologies for achieving this objective. We identify four challenges that pose high access barriers for the effective use of knowledge graphs, as their creation requires substantial knowledge of semantics and data modelling. Since their construction is a modelling task and since every model serves a purpose against which it is optimized, we question the central modelling paradigm of representing a mind-independent reality. Instead, we propose the Rosetta Statement approach, which models English natural language statements. We suggest a metamodel for Rosetta Statements, from which semantic data schema patterns for any type of simple English statement can be derived. We provide a simple light version and a full version that supports versioning and tracks changes for each Rosetta Statement. Rosetta Statements can be displayed in a user interface as natural language sentences. We implemented the Rosetta approach in the Open Research Knowledge Graph (ORKG), an open, domain agnostic, and community-driven knowledge graph for documenting research findings from scholarly publications. The ORKG enables domain experts without knowledge of semantics to define semantic data schema patterns for new types of Rosetta Statements. We discuss how the Rosetta Statement approach and our future plans solves the four challenges, also by for instance utilizing the structural proximity of Rosetta Statement patterns to natural language statements to develop supporting tools for entering data using Large Language Models. We discuss how the Rosetta approach supports a three-step procedure for FAIR knowledge graph construction, with the first step only requiring domain experts to use Rosetta Statements and Wikidata terms to create a FAIR knowledge graph that supports basic findability. In a subsequent step, semantic search capability can be added by replacing the Wikidata terms with ontology terms. Finally, if reason-capability is required, the graph can be updated by replacing Rosetta patterns with reasoning-capable schemata, which requires support from ontology engineers. This three-step procedure substantially lowers the entry barrier for knowledge graph construction and increases their cognitive interoperability.

# Introduction

We are experiencing an exponential increase in both data generation and consumption, with the aggregate data volume doubling every three years (1). Concurrently, the scholarly domain is witnessing a significant rise in publications, with an annual output exceeding 7 million academic papers (2). These figures underscore the urgency of harnessing **machine support**, as the sheer volume of data, information, and knowledge, without the assistance of machines, poses a threat to overwhelm and impede the acquisition of meaningful insights and fact-based decision-making.

The majority of research data are generated within projects, each with its own objectives, and are subsequently stored in project-specific databases or general repositories. The machine-actionability of these data is typically confined to a set of operations required by the project's objectives, resulting in datasets that are interoperable only within the context of the specific project and its operations in question. Consequently, each project-specific database or dataset in a repository has a tendency to become a data silo. With each project applying their own data structures and terminologies, there is a high probability that these data silos will **not be interoperable across projects**.

Considering that major global challenges, including biodiversity loss, zoonotic diseases, and climate change (3), require a truly interdisciplinary approach (4), where data and metadata must be collected, integrated, and analyzed from various sources and research fields, often involving the extraction of data from legacy literature, data tables on local hard drives, and relational databases, efficient machine-support can only be provided if both data and metadata are **machine-actionable** and **FAIR**, i.e., if they are readily **F**indable, **A**ccessible, **I**nteroperable, and **R**eusable for machines and humans alike (see the FAIR Guiding Principles (5)).

Unfortunately, conventional data management methodologies and techniques often encounter challenges in effectively handling the increasing volume, velocity, variety, and complexity of research data, making it challenging to retrieve, store, manage, handle, process, integrate, and analyze (meta)data efficiently within a reasonable timeframe (6). Moreover, these conventional methodologies and techniques lack the conceptual and technical requirements to efficiently support the generation of machine-actionable and FAIR (meta)data. Based on their transparent semantics, highly structured syntax, and standardized formats (7,8), semantic technologies such as knowledge graphs, semantic graph patterns (i.e., semantic models), and ontologies hold significant promise in addressing these challenges, facilitating the creation of machine-actionable FAIR (meta)data.

An **ontology** is composed of a set of resources representing classes (i.e., types of entities) and properties (i.e., relations and attributes) with commonly accepted definitions, aiming to provide a lexical or taxonomic framework for knowledge representation, with each resource being a Globally Unique Persistent and Resolvable Identifier (GUPRI) (9). Ontologies are used like dictionaries for creating formal, machine-actionable representations of a specific reality, formulated in a highly formalized, canonical syntax and standardized format, such as the Web Ontology Language[1] (OWL) that is based on description logic as formal logical framework and that can be serialized to the Resource Description Framework[2] (RDF). Ontologies generally comprise knowledge concerning types of entities pertinent to a specific domain, articulated through class axioms (terminology box; TBox) that delineate the attributes and relations to other types of entities that are inherent to every instance of the class. In essence, ontologies embody universal statements such as "*Every swan is white*" for defining a class 'swan', implying that if an entity is a swan, its color is necessarily white.

[1] https://www.w3.org/OWL/

[2] https://www.w3.org/RDF/

Assertional statements (e.g., "*Swan Anton is white*"), contingent statements (e.g., "*Swans can be white*"), and prototypical statements (e.g., "*Swans are typically white*" or "*Most swans are white*") establish relationships between instances and thus between individuals (assertion box; ABox). It is noteworthy that, contrary to universal statements, these types of statements are typically not covered in ontologies (10), yet they can be represented in **knowledge graphs** using the GUPRIs of respective ontology resources[3]. We therefore understand knowledge graphs to consist of a combination of empirical data in the form of ABox expressions and general domain knowledge in the form of TBox expressions, and distinguish them from ontologies, which primarily contain general domain knowledge in the form of TBox expressions and lexical statements (i.e., statements about linguistic entities, such as synonyms or preferred labels for a given term), but not empirical data.

The use of knowledge graphs, ontologies, OWL, and RDF alone does not guarantee compliance with the FAIR principles and does not automatically result in FAIR knowledge graphs with interoperable terms and machine-actionable and interoperable statements. The same ontology class or property must be used for referring to the same type of entity across different knowledge graphs to guarantee their **terminological interoperability**. For example, when referring to apples in different statements in a knowledge graph, the same ontology class should be used (e.g., apple (NCIT:C71985)). The same applies for the interoperability of statements. For a given type of statement, the same semantic graph pattern must be used for representing it in a knowledge graph to guarantee their **propositional interoperability** (for a discussion of machine-actionability, semantic interoperability, and the need for additional criteria for FAIR, see (12)).

A **semantic graph pattern** is a semantic model that describes relations between entities in a graph using resources and following the RDF syntax of *Subject-Predicate-Object*. In ontologies, semantic graph patterns take the form of **ontology design patterns** and are used for describing the relations between entities within TBox expressions. In knowledge graphs, they take the form of **semantic data schema patterns**, which are used for describing the relations between entities within ABox expressions. Tools for describing semantic graph patterns that enforce a standardized way of modelling and representing data of the same type exist, such as the Shapes Constraint Language SHACL[4] and Data Shapes DASH[5] (13), Shape Expressions ShEx[6] (14,15), or the Reasonable Ontology Templates OTTR[7] (16,17).

The effective employment of well-structured ontologies, FAIR knowledge graphs, and adequate semantic graph patterns has the potential to substantially increase the machine-actionability of (meta)data. However, it is crucial to note that (meta)data must also be human-actionable. We posit that data structures should be easily comprehensible for domain experts to support them in correctly interpreting and reusing them. Moreover, understanding the underlying semantic data schema patterns is key to writing queries for efficiently finding all relevant data in a knowledge graph. Therefore, we argued that interoperability, in essence, entails facilitating reliable exchange of information among machines and between humans and machines (12). The **Interoperability Framework** of the [European Open Science Cloud](European Open Science Cloud) (EOSC) differentiates technical, semantic, organizational, and legal interoperability as four discrete layers of interoperability for scientific data management (18). We proposed the incorporation of **cognitive interoperability**, as characterized in Box 1, as the fifth layer of interoperability within the EOSC Interoperability Framework (19).

---

[3] Although OWL, being based on description logic, only provides formal semantics for assertional statements as ABox expressions and for universal statements as TBox expressions, universal, contingent, and prototypical statements can be formally represented as ABox expressions using RDF (11).

[4] https://www.w3.org/TR/shacl/

[5] https://datashapes.org/forms.html

[6] https://shex.io/

[7] https://ottr.xyz/

**Box 1 | Cognitive Interoperability** (19)

Cognitive interoperability is "a critical characteristic of data structures and information technology systems that plays an essential role in facilitating efficient communication of data and metadata with human users. By providing intuitive tools and functions, systems that support cognitive interoperability enable users to gain an overview of data, locate data they are interested in, and explore related data points in semantically meaningful and intuitive ways. The concept of cognitive interoperability encompasses not only how humans prefer to interact with technology, i.e., **human-computer interaction**, but also how they interact with information, i.e., **human information interaction**, considering their general cognitive conditions. In the context of information technology systems such as KGs, achieving cognitive interoperability necessitates tools that increase the user's awareness of the system's contents, that aid in understanding their meaning, support data and metadata communication, enhance content trustworthiness, facilitate integration into other workflows and software tools, and that clarify available actions and data operations. Additionally, cognitive interoperability also encompasses ease of implementation of data structures and their management for developers and operators of information technology systems. It thus addresses the specific data, tool, and service needs of the three main personas (20) identified for users of information management systems such as KGs, namely **information management system builders** (i.e., information architects, database admins), **data analysts** (i.e., researchers, data scientists, machine learning experts), and **data consumers** (i.e., stakeholders, end users, domain experts)" (p. 11-12) (19).

The notion of cognitive interoperability emphasizes the enhancement of the usability of (meta)data structures and knowledge management systems for human users and developers. This aspect has been to a certain degree disregarded, particularly within the domain of knowledge graphs and semantic technologies. Furthermore, cognitive interoperability also takes into account the typical communication patterns of humans and their cognitive limitations.

In this paper, we introduce the **Rosetta Statement**[8] **knowledge graph construction and semantic modelling approach** to increase the cognitive and semantic interoperability of content in open and closed knowledge graphs[9] that have a cross-domain scope (see Box 2 for a description of the conventions that we follow throughout this paper). The *problem statement* section identifies four challenges of knowledge graphs relating to cognitive interoperability, graph querying, semantic parsing, and dynamic knowledge graph construction.

In the *result* section, we introduce the notion of semantic parsing as a modelling activity and argue that assertional statements in the form of natural language statements are models that share structural similarities with data structures, and that formalized natural language statements can be compared to table structures of relational databases and to semantic data schema patterns of knowledge graphs. We argue that semantic parsing involves a choice between different possible modelling approaches, with each model serving a specific purpose and being optimized for a specific data use. Based on these findings, we develop the Rosetta Statement approach to knowledge graph construction. Its emphasis lies on a modelling paradigm that enables machine-interpretability of (meta)data, prioritizing their findability and their interoperability across Rosetta Statements over their reasoning capabilities. This prioritization opens up new avenues for modelling by shifting away from the semantic parsing paradigm frequently applied in science that focuses on modelling a mind-independent reality. Instead, the Rosetta Statement approach models the structure of simple English natural language statements to enhance efficient and reliable communication of information between machines and between humans and machines. We introduce a light and a full version of the Rosetta Statement metamodel, with the latter also supporting versioning of statements and tracking the detailed editing history for each Rosetta Statement.

[8] The term refers to the Rosetta Stone, an ancient slab of black basalt containing the same text in three scripts: Ancient Greek, Egyptian hieroglyphs, and Demotic. It was instrumental in deciphering Egyptian hieroglyphs, as it provided a key to understanding their meanings. The Rosetta Stone is since a symbol of the power of decipherment and cross-cultural understanding.

[9] "Open" refers to the concept of accessibility, where users can register and contribute content to the graph, in contrast to a "closed" system, which involves a designated group of users or an automated workflow exists that manages content addition to the graph.

In the *Rosetta Statement use case: Open Research Knowledge Graph* section, we describe the implementation of the Rosetta Statement approach in the Open Research Knowledge Graph (ORKG) as an example use case. Furthermore, we discuss some of the future plans for further integrating the Rosetta Statement approach within the ORKG, adding services that utilize Large Language Models (LLMs) to support users in adding and finding semantic content in the ORKG.

In the *discussion* section, we discuss the benefits and potential issues we anticipate with applying the Rosetta Statement approach to knowledge graphs in general, and how the Rosetta Statement approach to knowledge graph construction could lower the barrier for creating FAIR and reasoning-capable OWL-based knowledge graphs by providing a first step in a three-step procedure.

**Box 2 | Conventions**

Throughout the paper, we use the term *triple* to denote a *Subject-Predicate-Object* triple statement, and *statement* to refer to a natural language statement. Also, when we talk about *schemata*, we explicitly include schemata for statements and for collections of statements and not only schemata for individual triples.

When we use the term *resource*, we mean something that is uniquely designated, such as through a Uniform Resource Identifier (URI), and that serves as an object of discussion. A resource thus stands for something and represents something someone wants to talk about—it represents something of interest. In the context of RDF, both the *Subject* and the *Predicate* in a triple are always considered resources, while the *Object* can be a resource or a literal. Resources can represent properties, instances, or classes. Properties are used in the *Predicate* position, instances denote individuals (i.e., instances), and classes represent general categories, universals, or types.

To ensure clarity, both in the text and in all figures, we represent resources using human-readable labels instead of their URIs. It is implicitly assumed that each property, instance, and class possesses its own URI. All resources relating to Rosetta Statements use the prefix 'rosetta' (e.g., '*rosetta:rosetta statement*') and are defined in the Rosetta Statements Ontology[10].

# Problem statement

## Cognitive interoperability challenge: Understanding machine-actionable semantic data schema patterns requires knowledge and experience in semantics

Humans are experts in efficiently communicating information by omitting background knowledge, employing vague statements that allude to general figures of thought, and by utilizing metaphors and metonymies[11]. We are adept at minimizing the amount of information needed to be conveyed, relying on the context for others to infer the missing details.

[10] https://github.com/karanthpallavi/rosetta_statement_ontology

Contrary to humans, machines require explicit presentation of all relevant information, resulting in the challenge arising from the conflict between machine-actionability and human-actionability of (meta)data representations: **as data representations become more geared towards machine-actionability, they become more complex and less readily understandable for humans (i.e., less human-actionable)** (19).

The formal semantic representation of the statement "*This apple has a weight of 212.45 grams, with a 95% confidence interval of 212.44 to 212.47 grams*" in a knowledge graph as displayed in Figure 1 makes sense from a machine and data management perspective. It complies with the commonly applied modelling paradigm of truthfully representing the relations between real entities, thus, attempting to create a digital twin that models a mind-independent reality. The semantic representation also enables semantic reasoning over the graph. At the same time, it takes into account the need for a modular approach to structure data in a knowledge graph and the need for reusing semantic data schema patterns to reduce the variety and overall complexity of patterns used within a knowledge graph.

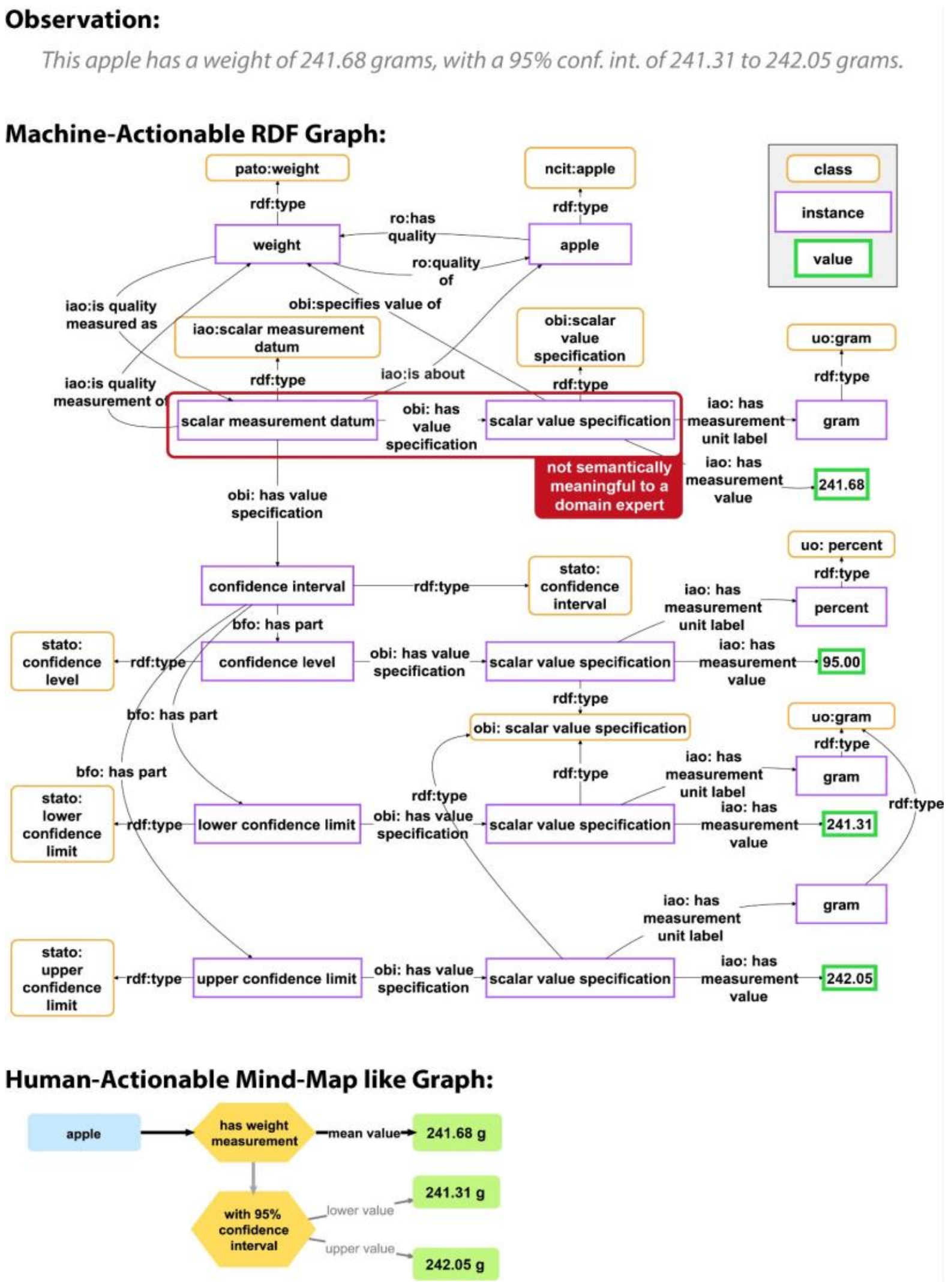

**Figure 1: Comparison of a human-readable statement with its machine-actionable and its human-actionable representation. Top**: A human-readable statement about the observation that a particular apple weighs 241.68 grams, with a 95% confidence interval of 241.31 to 242.05 grams. **Middle**: A machine-actionable representation of the same statement as an ABox semantic graph, using RDF syntax and following the general pattern for measurement data from the Ontology for Biomedical Investigations (OBI) (21) of the Open Biological and Biomedical Ontology (OBO) Foundry. Marked in red is a triple that is not semantically meaningful to a domain expert and thus difficult to comprehend for them. **Bottom**: A human-actionable representation of the same statement as a mind-map like graph, reducing the complexity of the RDF graph to the information that is actually relevant to a human reader.

[11] A metonymy is the substitution of the name of an attribute or adjunct for that of the thing meant, for example, *"the pen is mightier than the sword"* with *"the pen"* referring to writing and *"the sword"* to warfare and violence.

However, exploring such machine-actionable graphs and extracting the information that is essential for the underlying statement can become a challenge. Especially, if the user interface (UI) of the knowledge graph employs a **semantic data browser** for accessing information from the graph, which allows users to start exploring from a single resource as the entry point, and moving from here along triple paths following RDF links (22,23). For example, for extracting all relevant information of the apple measurement statement in Figure 1 using a semantic data browser, a user would have to click 15 times (19).

Moreover, from the perspective of most domain experts, the graph in Figure 1 is also overly complex, hard to understand, and includes information that is not relevant and often even incomprehensible to a human reader. In short: domain experts do not like to look at graphs like the one displayed in Figure 1. This impedance mismatch has the potential to frustrate humans when communicating (meta)data with machines.

If we want to store (meta)data in a knowledge graph in a machine-actionable format and simultaneously present them in an easily understandable, human-readable way in a UI, it is necessary to decouple the data storage in the graph from the data presentation in the UI, so that information that is necessary for machines but irrelevant for humans is only accessed by machines but not displayed in the UI (see Fig. 1, bottom). Alternatively, data structures that are easily comprehensible to a domain expert could be created.

## Graph query challenge: Writing queries for a knowledge graph requires knowledge of graph query languages

As a domain expert using a knowledge graph, it is one thing to comprehend a given (meta)data structure, and it is another thing to actually find (meta)data that interests you. **Findability** is the most important aspect of any tool that stores and documents (meta)data. If relevant (meta)data cannot be found in the first place, interoperability issues become secondary.

In the context of knowledge graphs, specific query endpoints can be used along with corresponding **graph query languages** to interact with the graph. Querying a knowledge graph thus requires, either directly or indirectly through a UI, writing queries with such a graph query language—SPARQL for RDF- and OWL-based knowledge graphs and Cypher for labeled property graphs such as Neo4J.

Our personal experience is that most users and software developers have no experience with graph-based databases, are not familiar with graph query languages and their benefits, and therefore do not see the need to learn them. And even those who are familiar with them report that writing more complex queries can be demanding and is time-consuming and error-prone, requiring knowledge about the underlying semantic data schema patterns used in the knowledge graph. **Domain experts are usually not familiar with graph query languages** and do not know how to write queries with them. They cannot take advantage of the full search-capabilities of a knowledge graph if no intermediate interface is used that translates a natural language question into a SPARQL or Cypher query. **Apparently, the need to write SPARQL or Cypher queries is a barrier to interacting with knowledge graphs and hinders their wider use** (24)).

Recent research suggests that this challenge can be circumvented using LLMs that translate natural language questions into SPARQL queries (25). However, using LLMs in this context can lead to misinterpretations due to the ambiguity present in natural language expressions, resulting in inaccurate and unintended outcomes (26). Moreover, LLMs seem to be less efficient when dealing with intricate and nested queries (27). It remains to be seen whether future developments and improvements of LLMs will be able to overcome these weaknesses.

## Semantic parsing burden challenge: Knowledge graph construction requires expertise and experience in semantics and semantic modelling

**Semantic parsing** is the task of translating a natural language utterance, a data structure from a relational database, or data from a CSV file into a machine-interpretable representation of its meaning in a knowledge graph. It typically involves the use of a formal language, such as OWL, and follows the triple syntax of *Subject—Predicate—Object*. As such, it is an essential part of constructing a semantic knowledge graph and is usually carried out by someone with experience in semantics and semantic data modelling. It involves the development of semantic data schema patterns and is a major challenge for rapidly building knowledge graphs with FAIR machine-actionable (meta)data. Especially, if the semantic parsing follows the paradigm that the output graph should represent a mind-independent reality that can be reasoned about.

Depending on the context and the complexity of the system-of-interest to be modelled, the development of semantic data schema patterns that truthfully represent the relationships between real entities can be very time-consuming and overall demanding, even for someone who is experienced in semantics and has done such modelling before. Unfortunately, the **typical domain expert** who produces the data to be parsed, and who therefore has the best understanding of the data, **is usually not an expert in semantics and semantic parsing**. They do not know how to model the data in terms of formal semantics using the *Subject—Predicate—Object* syntax. And they do not know how to create logically consistent semantic data schema patterns using existing ontology classes and properties.

Consequently, the development of such patterns necessitates close collaboration between domain experts and semantics experts, a process that is often time-consuming and not always feasible due to limited funding and a shortage of semantics experts. This results in a significant **semantic parsing burden**, which is particularly critical in the context of community-driven dynamic knowledge graph construction (see next *challenge*). While LLMs have shown considerable efficacy in semi-automating knowledge graph and ontology construction from input texts, they necessitate a human-in-the-loop for quality control and typically require the provision of semantic data schema patterns for their prompts (28–32). Consequently, they do not offer a solution to the semantic parsing burden.

## Dynamic knowledge graph construction challenge: knowledge graph construction and semantic interoperability

The overall expressive power of the *Subject—Predicate—Object* triple structure of a knowledge graph allows for a wide range of modelling possibilities for any given information, with the same information likely being modelled in numerous and fundamentally different ways across different knowledge graphs. If a knowledge graph does not restrict the modelling choices for a specific type of information to a single semantic data schema pattern and if it does not restrict the choice of ontology terms to be used in this pattern through semantic slot-constraints, substantial problems with semantic interoperability will arise that will affect terminological as well as propositional interoperability, ultimately impacting the findability, interoperability, and reusability of the information and thus the overall FAIRness of the (meta)data (12).

Most knowledge graphs follow the abovementioned modelling paradigm of truthfully representing a mind-independent reality. They typically focus on a specific scope and restrict their content to a fixed set of different types of information. For each type of information, they pre-define a corresponding semantic data schema pattern together with a specification of its semantic slot-constraints. Respective knowledge graphs employ a **static knowledge graph construction approach** based on static information extraction. The restriction on the use of only one semantic data schema pattern for each type of information and the specification of semantic slot-constraints for each pattern ensures the logical consistency and semantic interoperability of the graph's content, resulting in a truly FAIR knowledge graph that supports reasoning. By closely collaborating with domain experts, it is a feasible task for ontology engineers to define a limited set of semantic data schema patterns and semantic slot-constraints that are required for such a knowledge graph.

With their **open**, **domain agnostic scope**, **knowledge graphs** such as Wikidata or the ORKG (33,34), however, cannot adhere to this conventional static information extraction approach. Instead, they follow a **community-driven dynamic knowledge graph construction (DKGC) approach**, where the graph's coverage of different types of information is continuously evolving through the input of their users. Knowledge graphs following the DKGC approach face unique challenges. It is not feasible to pre-define all semantic data schema patterns and ontology terms required for modelling all possible types of information users may want to add to the graph. As a result, users must handle semantic parsing themselves, and usually without the support of any ontology engineer. This creates a significant barrier to data entry, likely leads to semantic ambiguities, logical inconsistencies across the graph, general data quality issues, and ultimately a lack of semantic interoperability and FAIRness of information in the graph, all of which limits the findability of information within the graph (see *Semantic parsing burden challenge*). Moreover, due to their community-driven data entry procedures, DKGC approaches usually require versioning of the graph, as users can make modifications at any time. Ideally, the versioning includes a detailed editing history to ensure transparency and build trust.

A further consequence of the DKGC approach in domain agnostic knowledge graphs is the practical and theoretical impossibility of supporting reasoning over the entire graph. From a practical standpoint, it is impossible to pre-define all the semantic data schema patterns and accompanying ontology terms that are required to model information across all possible domains. From a theoretical standpoint, it is impossible to create ontologies that are both logically consistent with each other and tailored to meet the unique needs of each domain. To illustrate this challenge, consider the disparity between Newtonian physics, which asserts that an electron is a mass particle but not a wave, and quantum physics, which recognizes the electron as both a mass particle and a wave. Both ontological frameworks are crucial to physicists, and the choice between them depends on the specific experiment to be documented and the overall scope of the study. A similarly impossible challenge is the development of semantic data schema patterns that are optimized for all possible operations, as this would require addressing the diverse usage needs of the various users of the knowledge graph in a single semantic data schema pattern (12).

# Result

We believe that the implementation of the Rosetta Statement approach to semantic modelling and knowledge graph construction, a modelling paradigm that models natural language statements as opposed to attempting to realistically represent a mind-independent reality, has the potential to make a substantial contribution to the resolution of the four aforementioned challenges. However, prior to the introduction of the Rosetta Statement approach, we first discuss semantic parsing as a modelling approach.

## Semantic parsing—a modelling approach

In general, a model is defined as a representation of information on something (i.e., meaning), created by a sender for a receiver, with a specific purpose and usage context in mind (35). The model's purpose is to act as a surrogate for the system-of-interest that it represents; its responses should be consistent with those of the actual system, however, focusing only on the properties relevant to its intended use (36). For a model to be effective, it must possess the following three features (37):

1. **Mapping feature**: The model is derived from and attempts to represent a system-of-interest.
2. **Reduction feature**: The model includes only a relevant subset of the system's properties; abstraction is essential for modelling.
3. **Pragmatic feature**: The model must be usable as a substitute for the system-of-interest in relation to its specific purpose.

With this understanding, both the structures underlying assertional natural language statements and data structures can be seen as models (12). Models can be differentiated into token and type models (36).

### Assertional statements and empirical data are token models

A **token model** (also known as *snapshot model*, *representation model*, or *instance model*) captures specific properties of elements from the system it represents, maintaining a one-to-one correspondence with the system. It reflects individual attribute values, such as the weight of a particular apple. **Token models therefore represent the relationships between individual entities** (i.e., instances) belonging to the modelled system. The creation of a token model involves selecting which properties to include (**projection**) and converting these properties into elements of the model (**translation**). Elements within a token model align with and correspond to specific elements of the modelled system-of-interest—for instance, a particular apple and its weight. Consequently, different token models of the same system-of-interest, representing the same set of properties, are connected through a transitive token-model-of relationship. This relationship can be organized into sequences of designators, each sequentially representing its corresponding element across all token models, ultimately pointing back to the original element in the system-of-interest (e.g., 'apple' in model *C* to 'apple' in model *B* to 'apple' in model *A* to the real apple from the system-of-interest) (36).

According to this definition, assertional statements in the form of natural language sentences and empirical data both can be understood as token models (12). Figure 2 shows examples of different (types of) token models of the same system-of-interest, including the sentence at the top (Fig. 2A) and the tabular and graph-based data structures at the bottom (Fig. 2D, E).

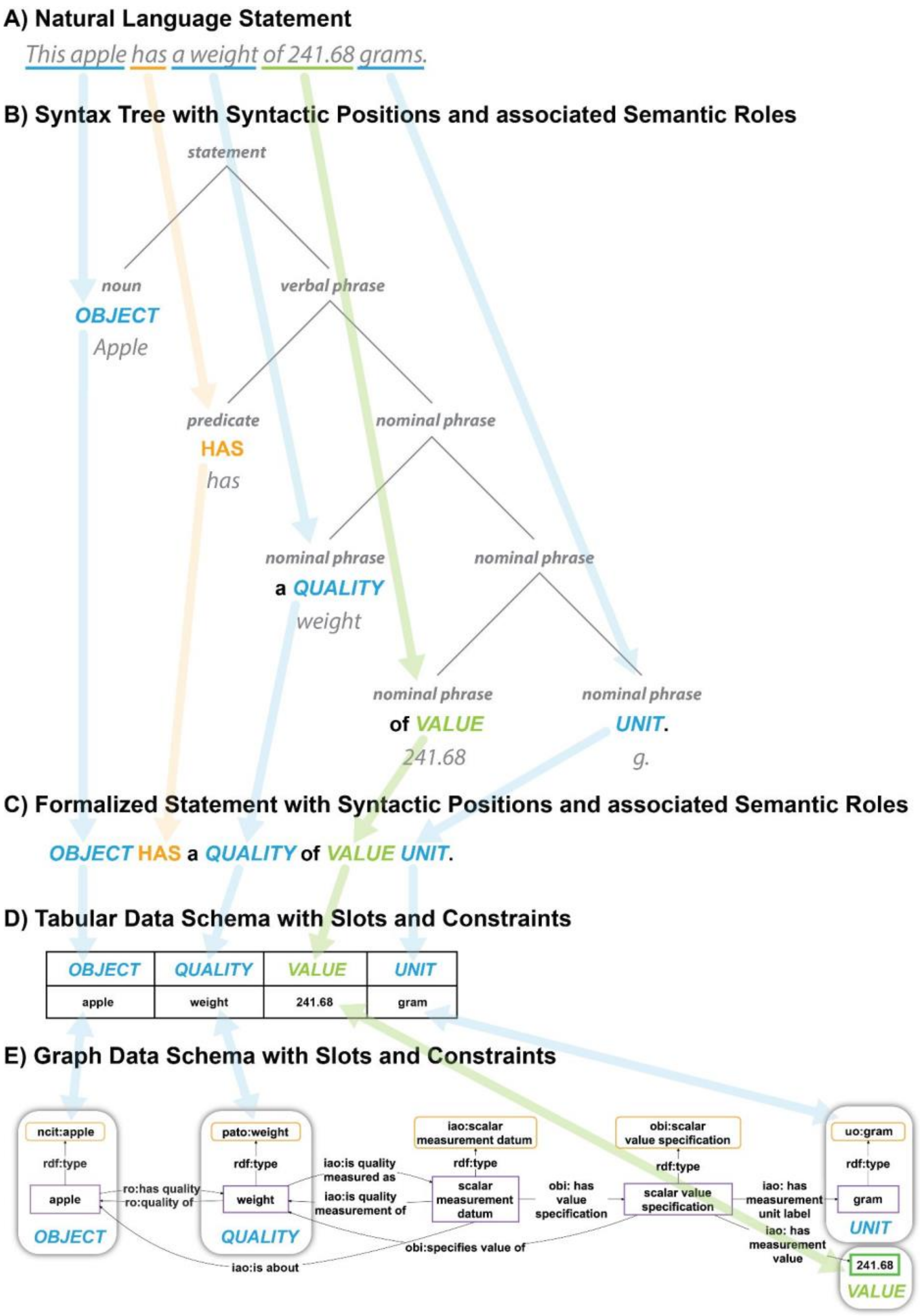

**Figure 2: Parallels between natural language statements and data schemata. A)** A natural language statement is structured by syntactic and grammatical conventions into syntactic positions of phrases of a syntax tree. **B)** The syntax tree corresponding to the natural language statement from A). **C)** The formalized statement of the natural language statement from A), where each position is associated with a specific semantic role, which can be described by a thematic label. **D)** A tabular data schema of the natural language statement from A). **E)** A graphical data schema of the natural language statement from A). Both data schemata must represent the syntactic positions of the natural language statement as slots, and each slot must specify its associated semantic role as a constraint specification.

Natural language token models differ from data structure token models in their specific purpose. While the former primarily serve the purpose of communicating information about the system-of-interest between humans, the latter serve the purpose of communicating information between machines and additionally serve analytical purposes.

Understanding natural language statements as token models aligns with the **predicate-argument-structure** framework in linguistics (38,39), where the main verb of a statement and its auxiliaries form the predicate. The predicate's **valence** specifies the number and types of subjects and objects needed to complete its meaning (called **arguments**). Additional objects (called **adjuncts**) that provide optional information, such as a time specification in a parthood statement, may also be related to the predicate. Every statement, thus, includes a subject phrase as one of its arguments and can have, depending on the underlying predicate, one or more object phrases as further arguments and additional adjuncts.

In the syntactic structure of a statement, each argument and adjunct occupies a specific **syntactic position**, with each position having its own **semantic role** (Fig. 2B; see also Kipper et al.'s (40) verb lexicon VerbNet, which extends Levin verb classes (41); see also *thematic roles* sensu (40)). Each position can be described using a **thematic label** that reflects the position's semantic role (e.g., OBJECT, QUALITY, VALUE, UNIT in Fig. 2). The syntactic structure of a given statement can then be represented as a **syntactic frame** of a sequential order of thematic labels forming a formalized statement (Fig. 2C) ((40), see also PropBank (42)).

Assertional natural language statements such as "*This apple has a weight of 241.68 grams*" can be understood to be token models, with the apple's weight property modelled via corresponding syntactic positions. Each (tabular and graph-based) data model of the same apple modelling the same property via corresponding slots is also a token model, and all these token models relate to each other via a transitive token-model-of relationship. Consequently, we can understand each empirical datum as the formalized representation of the same system-of-interest as is represented in the corresponding assertional natural language statement, where **slots of the data structure** can be **aligned** with and compared to **syntactical positions**, with the semantic constraints of a slot mapping to the position's associated semantic role (12). The main difference between these models is their purpose, with natural language statements being used for human communication whereas data structures are designed to be easily read and operationalized by machines.

## Formalized assertional statements, table structures, and semantic data schema patterns are type models

A **type model** (also known as *schema model* or *universal model*) can be derived from a token model by **classification** of its properties. The formalized statement in Figure 2C, for example, is the type model of the corresponding natural language token model (Fig. 2A) and can be obtained from the latter via the corresponding syntax tree (Fig. 2B) by classifying the individual entities in the subject and object positions (e.g., the individual entity "*this apple*" to an instance of the class '*apple*'). By further generalizing the identified classes, one can then obtain the semantic role of each position (e.g., the semantic role *OBJECT* from the class '*apple*'), resulting in a metamodel. A **metamodel** is a model of a model that can be obtained by **generalizing** over a given type model (36). Metamodels represent a specific kind of type model and are more broadly applicable. By generalization, the formal statement type model "*APPLE HAS a WEIGHT of VALUE GRAM-BASED-UNIT*" can be transformed into the formal statement metamodel "*OBJECT HAS a QUALITY of VALUE UNIT*" (Fig. 2C).

The structures used for organizing a datum as a row in a table (Fig. 2D) or a subgraph in a knowledge graph (Fig. 2E) are metamodels as well, and correspond with their related formal statement metamodel (12). The constraint for a column in a data table or a slot in a data graph specifies an ontology class that defines which instances are allowed as input and aligns with the semantic role of the corresponding syntactic position.

A given datum is thus a token model that is typically created by instantiating a corresponding data schema that is its metamodel (12). The dependency of data token models from their underlying data schema metamodels serves the purpose of supporting machine-actionability and semantic interoperability across data of the same type.

In terms of cognitive interoperability, we can conclude that a data schema metamodel (e.g., a semantic data schema pattern) must provide a structure that is functionally and semantically similar to a formalized statement, involving the same elements as the syntax tree of the corresponding natural language token model, in order to be intelligible to a human reader. The metamodel must thus cover all relevant syntactic positions as slots, with their associated semantic roles modelled as constraint specifications. Only if this **minimum requirement** is met, humans will be able to understand data created based on a metamodel by translating it into a corresponding natural language statement (12). In the process of creating data metamodels, such as tables in a relational database or semantic data schema patterns for a knowledge graph, it is imperative to comprehend them as attempts to translate the structure of natural language statements into machine-actionable data structures. This is due to the fact that human readers require the ability to effortlessly translate a datum back into its corresponding natural language statement to truly comprehend the information contained in it. In the event that this minimum requirement is not met, human readers are likely to misinterpret the data structure.

## Semantic parsing: Choosing between different type models

In the process of modelling a system-of-interest to capture specific aspects of reality, we create **representational artifacts** and thus entities that carry meaning and that we use for communicating about that reality (43,44). Two kinds of representational artifacts can be distinguished. **Iconic representational artifacts** carry perceptual non-conceptual content such as images, videos, 3D models, physical objects in a collection, audio recordings, or diagrams. In these cases, meaning is contained via a natural relation of resemblance to the part of reality that it reproduces (*natural meaning* (45,46)). In contrast, **textual representational artifacts** carry semantic conceptual content by using words and symbols, which, in turn, convey meaning based on common agreement (*non-natural meaning* (45)).

Notably, only semantic content and thus textual representational artifacts can be analyzed and processed by a computer[12], and only to them, logical reasoning can be applied. However, both types of representational artifacts serve as models of reality that play an essential role in scientific communication. In this paper, we focus on textual representational artifacts and their representation in a knowledge graph[13].

For representing a given semantic content, there usually exist many possible natural language token models. The content of the sentence "*This apple has a weight of 241.68 grams*" could have equally been modelled as "*The weight of this apple is 241.68 grams*" or "*241.68 grams is the weight of this apple*". The same applies to data structures, both tabular and graph-based (e.g., Fig. 2D,E and Fig. 3).

In the context of knowledge graph construction, the semantic parsing task thus involves the choice between all possible models, with the goal to ideally apply only one semantic data schema pattern for representing a given type of data. Models, however, are typically designed with a specific purpose and usage context in mind, against which they are optimized. Consequently, the model choice should be based on the purpose and thus the anticipated usage of the content in the knowledge graph, since no data schema can be optimal across all different usage contexts but is always context and format dependent (12). However, if more than one usage is anticipated, it is very likely that more than one data schema must be used for modelling them, resulting in schema interoperability issues. We can deal with this in a knowledge graph and establish schema interoperability by defining schema crosswalks between semantic data schema patterns that model the same semantic content and thus the same system-of-interest (see Fig. 3) (12).

---

[12] Perceptual content must be translated into semantic content before a computer can analyze it.

[13] Attempts to represent perceptual contents in a knowledge graph exist (i.e., multi-modal knowledge graphs; e.g., (47,48)), but are based on semantically annotating bearers of perceptual content so that they can be integrated with the semantic content of the knowledge graph.

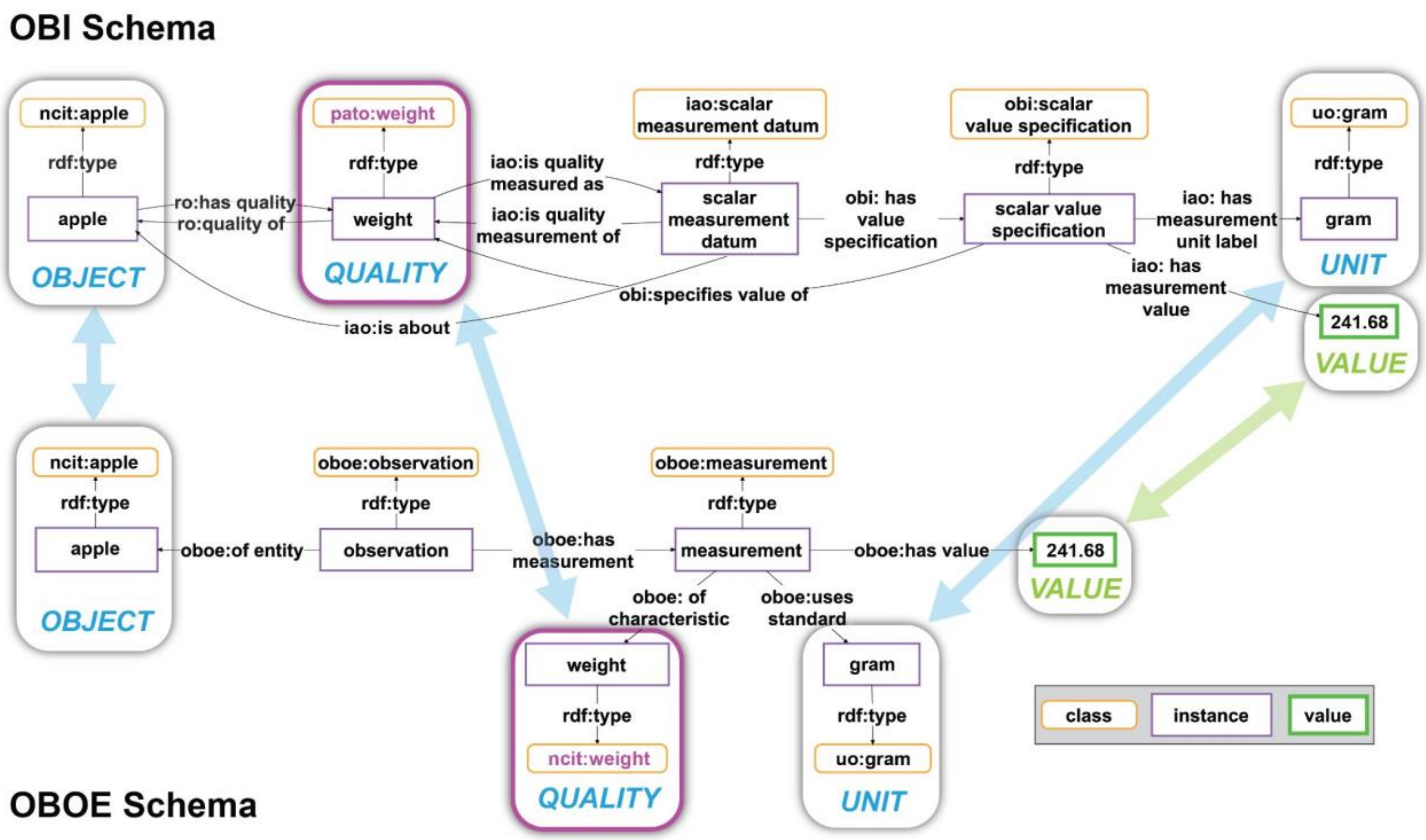

**Figure 3: Crosswalk from one schema to another for a weight measurement statement.** The same weight measurement statement is modeled using two different schemata. **Top**: The weight measurement according to the schema of the Ontology for Biomedical Investigations (OBI) (21) of the Open Biological and Biomedical Ontology (OBO) Foundry, which is often used in the biomedical domain. **Bottom**: The same weight measurement according to the schema of the Extensible Observation Ontology (OBOE), which is often used in the ecology community. The arrows indicate the alignment of slots that share the same constraint specification, i.e., the same semantic role. The corresponding semantic roles include the *OBJECT*, the *QUALITY*, and the *VALUE* that has been measured together with its *UNIT*. The slots and their relationships to one another carry the semantic content that actually conveys the meaning of the weight measurement statement to a human reader. Blue arrows indicate slots with resources as values, and green arrows those with values. Slots with purple borders indicate issues with terminological interoperability: OBO uses an instance of the class ‘pato:weight’, while OBOE, in this example, uses an instance of the class ‘ncit:weight’. However, since ‘pato:weight’ and ‘ncit:weight’ are synonymous terms and can therefore be mapped to establish terminological interoperability between them, the two graphs are semantically interoperable.

In light of the unpredictability of the purposes and potential uses that users of a knowledge graph might have for its semantic content—uses which often depend on a particular research question and the tools and methods used for data analysis—it is not feasible to provide semantic data schema patterns that support all possible uses and purposes. However, two main domain agnostic purposes can be identified that apply to all knowledge graphs: reasoning and FAIRness.

**Reasoning** is employed in a knowledge graph to assess its logical consistency, to automatically classify instances within the graph, and to infer implicit knowledge derived from TBox expressions (e.g., class axioms, logical characteristics of a given property). It is applied to the graph to complement it by adding ABox expressions in the form of additional triples. Reasoning is based on a logical framework. In knowledge graphs, this framework is typically **description logic**, necessitating data to be semantically modelled using a formal language such as OWL. When designing semantic data schema patterns that support reasoning, the semantic parsing paradigm of representing a mind-independent reality is generally applied. However, this paradigm is associated with challenges, including issues with cognitive interoperability arising from overly complex and incomprehensible graphs (see *Cognitive interoperability challenge*), as well as a high resource intensity and the necessity for collaboration between domain experts and semantics experts, who are often in short supply (see *Semantic parsing burden challenge*). Additionally, with an increasing size and interconnectivity of the graph, maintaining the logical consistency of the semantic content in the graph becomes increasingly difficult. In essence, the modelling of a mind-independent reality for the purpose of reasoning imposes a substantial barrier on the semantic parsing task, which has a negative effect on the overall acceptance of knowledge graphs, particularly in the context of documenting information derived from and knowledge gained in smaller research projects. Moreover, as previously discussed in the *dynamic knowledge graph construction challenge*, community-driven and open domain agnostic knowledge graphs cannot adopt this semantic parsing paradigm due to practical and theoretical limitations, which consequently limits their use to purposes other than reasoning.

While supporting reasoning is a valuable purpose for a research knowledge graph, other purposes can be valuable as well. We posit that supporting **FAIRness of (meta)data with high cognitive interoperability** is of even greater importance for research knowledge graphs. The ability to find relevant data is a prerequisite for using them, and for all other data usage, including reasoning, adequate semantic data schema patterns with corresponding schema crosswalks can always be defined as needed. However, if reasoning is not the primary objective of modelling information and knowledge in a knowledge graph, semantic parsing can be liberated from this meticulous requirement, and other pathways for representing semantic content can be investigated.

## The Rosetta Statement approach to semantic parsing

With the Rosetta Statement approach, we specify a metamodel in the form of a general semantic data schema pattern that serves the specific purpose and usage context of **facilitating communication of semantic content between machines and domain experts**. The resulting representations are expected to adhere to all criteria of the FAIR Principles and to ensure high cognitive interoperability that manifests itself in representations of semantic content in the graph that are readily comprehensible to domain experts.

Humans typically communicate semantic content through natural language expressions. In light of this, we propose that the general semantic data schema pattern be modelled after the structure of natural language statements. The fundamental premise underlying the Rosetta Statement approach is, therefore, to model simple natural language statements.

A Rosetta Statement represents a **smallest semantic-content-carrying unit of information that is semantically meaningful to a human reader** (cf. *statement unit* in (49); for an example of triples in the graph that are not semantically meaningful, see Fig. 1 middle). The main purpose of this approach to semantic parsing is to support the communication of semantic content in a knowledge graph with domain experts and to **reduce the burden of semantic parsing** and thus knowledge graph construction and with it to lower the barrier for the use of knowledge graphs by domain experts.

While terms carry meaning through their ontological definitions, statements carry meaning through their terms and the syntactic positions in which they are placed. Unfortunately, when looking at the predicate-argument-structure and comparing the structure of triples with that of natural language statements, we see that they are quite different and therefore do not properly align: the *Predicate* of a triple is always and necessarily **binary**, i.e., triples always have exactly one subject and one object argument. This is not the case for natural language statements. Although binary natural language predicates exist, as for example in the statement “*This tree* carries *an apple*”, not every natural language predicate is necessarily binary. The statements "*This apple* has *a weight* of *212.45 grams*" and "*Anna* travels by *train* from *Berlin* to *Paris* on the *21st of April 2023*" provide examples of statements with **n-ary** predicates. Therefore, ontology properties (i.e., the *Predicate* resources used in triples) do not map in a one-to-one relation to natural language predicates, and we often need to use multiple triples to model a natural language statement (cf. Fig. 1 and Fig. 2E). Modelling n-ary statements using the *Subject—Predicate—Object* triple structure is possible in principle, but requires the construction of a subgraph consisting of multiple triples.

In the following, we introduce a modelling paradigm that reflects the structure of English natural language statements and provides a generic pattern for modelling n-ary statements in RDF that functions as a metamodel. In all of this, we try to take a **pragmatic approach** that may not satisfy all the requirements for knowledge management one would wish for in an ideal world, but which we hope will bring practical improvements in overall usability and comprehensibility for all users of knowledge graphs.

## Requirements for a Rosetta Statement metamodel

At the core of the Rosetta Statement approach is a distinct **modelling paradigm for statement types**. In order to satisfy the requirements of cognitive interoperability, it is essential to employ a modelling paradigm that is as generic and simple as possible, reflecting as much as possible the structures with which we are already familiar from natural languages such as English. In addition, the paradigm must enable the specification of new Rosetta Statement schemata, thereby facilitating a streamlined process that does not necessitate a background in semantics—the approach should allow for **the automation of the semantic parsing step** to the greatest extent possible.

Achieving this objective necessitates the formulation of a **highly generic metamodel structure**, one that is applicable to any statement type, irrespective of its **n-arity**. To ensure efficiency, models derived from this metamodel should comprise solely the information necessary to recuperate the natural language statement's meaning. This information should be the equivalent to that required to generate a new natural language statement of the corresponding type from a user.

For example, instead of creating the entire subgraph as shown in Figure 2E, for a weight measurement statement it should be sufficient to store only the resources for the measured object and the quality, together with the value and the unit, with the emphasis on **always being able to reconstruct the original user input or data import for a given statement by storing only the semantically constitutive entities**, i.e., those objects and relations that preserve the core semantics of the natural language statement, and thus those entities that align with all syntactic positions required to translate the semantic content into a natural language statement.

Another requirement that the generic metamodel must satisfy is that it must facilitate the seamless derivation of queries from itself and from any of the statement-type specific semantic data schema patterns derived from it (see *graph query challenge*).

Each of these requirements is important because, ultimately, the metamodel and all semantic data schema patterns derived from it must support semantically interoperable (meta)data statements with which not only machines but also humans can interact. The question, then, is how to achieve this objective?

It is imperative to abstract the structure of syntax trees from natural language statements to their syntactic positions and associated semantic roles. Given that full expressiveness of natural language statements is not required when documenting (meta)data statements, it is sufficient to model statements with a relatively simple structure, comprising a subject, a transitive verb or predicate, and a number of objects. In this first attempt to formulate machine-actionable Rosetta Statements, we do not consider passive forms and tenses, while also abstaining from distinguishing between various syntactic alternations in which a verb or predicate can express its arguments. **The metamodel underlying our modelling paradigm is, therefore, analogous to a highly simplified syntactic frame, i.e., a formalized statement** (see Fig. 2 C), specifying a subject-position and a number of required and optional object-positions, each with its associated semantic role characterized as thematic label and a corresponding constraint specification. The structure of the metamodel is, thus, an abstraction of the structure of a syntax tree.

Different types of Rosetta Statements can be distinguished on the basis of their underlying predicates (i.e., relations). This results in a **predicate-based classification** of types of Rosetta Statements[14]. For example, the statement about the weight of an apple (Fig. 2) is a Rosetta Statement of the type *measurement statement*.

Statements also differ in their number of objects. A statement such as '*Sarah* met *Bob*' exemplifies a **binary relation**, where '*Sarah*' is designated as the subject and '*Bob*' as the object. The addition of a date, such as '*Sarah* met *Bob* on *4th of July 2021*', transforms the statement into a **ternary relation**, comprising two objects[15]. The addition of a place, for example, transforms the relation into a **quaternary relation**, as in '*Sarah* met *Bob* on *4th of July 2021* in *New York City*'. This is an open-ended relation in principle, although its extent is limited by the dimensionality of the human reader's ability to comprehend n-ary relations[16]. Notwithstanding this limitation, statements can be distinguished based on their n-arity.

Furthermore, a distinction can be made between arguments and adjuncts, enabling the differentiation of **objects necessary** for the completion of the meaning of the statement's predicate from **objects** that are **optional**.

Moreover, if we were to model the statement '*Sarah* met *Bob* on *4th of July 2021*' in a knowledge graph, the objects '*Bob*' and '*4th of July 2021*' would be modeled differently. Whereas '*Bob*' is likely to be modeled as a resource that instantiates a class 'person' (wikidata:Q215627), '*4th of July 2021*' is likely to be modeled as a literal associated with the datatype xsd:date. Consequently, in addition to differentiating arguments and adjuncts, each with their associated semantic roles and thematic labels, one can distinguish objects by their type into resources (via their respective GUPRIs) and literals. **Ontology resources** function as constraints for resources and **datatypes** for literals, both types of **constraints aligning with the associated semantic roles**. Resources, in turn, can be either named-individuals, classes, or properties. The former will be referred to as **resource-objects** and the latter as **literal-objects**.

After having identified the different subject and object positions within a statement, the next step is to classify and generalize each position to identify its semantic role and specify its constraints. This results in the specification of a formalized statement and, consequently, a natural language metamodel that can be translated into a Rosetta Statement schema pattern. To illustrate, the statement '*Sarah* met *Bob* on *4th of July 2021* in *New York City*' transforms into the natural language metamodel '*PERSON* met *PERSON* on *DATE* in *LOCATION*'.

[14] This classification could also be based on different syntactic frames (see (40,42)).

[15] Many properties of the Basic Formal Ontology2020 are actually ternary relations because they are time-dependent (50,51). For example, "*subject* located in *object_A* at *t*".

[16] Humans can hold only 5–9 items in memory (52).

With the introduction of the concept of resource-subjects, resource-objects, and literal-objects, we now possess the various elements that each Rosetta Statement schema pattern must encompass. The subsequent step involves the optimal arrangement of these elements to each other and to the statement resource.

## The light version of the Rosetta Statement metamodel

The Rosetta Statement modelling approach requires relating the subject-resource of a given statement to all of its distinct object-resources and object-literals. To circumvent challenges that commonly emerge when modelling statements that possess n-ary predicates and to closely mirror the structure of simple natural language statements in English—statements comprising only one verb or predicate—a Rosetta Statement ontology class is defined for each statement type based on the statement's verb/predicate. An instance of the respective class is used when creating a new statement, linking the subject and all object-resources as well as object-literals to it (Fig. 4). For instance, the statement '*This apple has a weight of 241.68 grams*' would instantiate a '*has-measurement statement*' Rosetta Statement class. The corresponding semantic data schema pattern would link an instance of '*apple*' (wikidata:Q89)[17] as the statement's subject-resource via a '*subject*' property to an instance of this class.

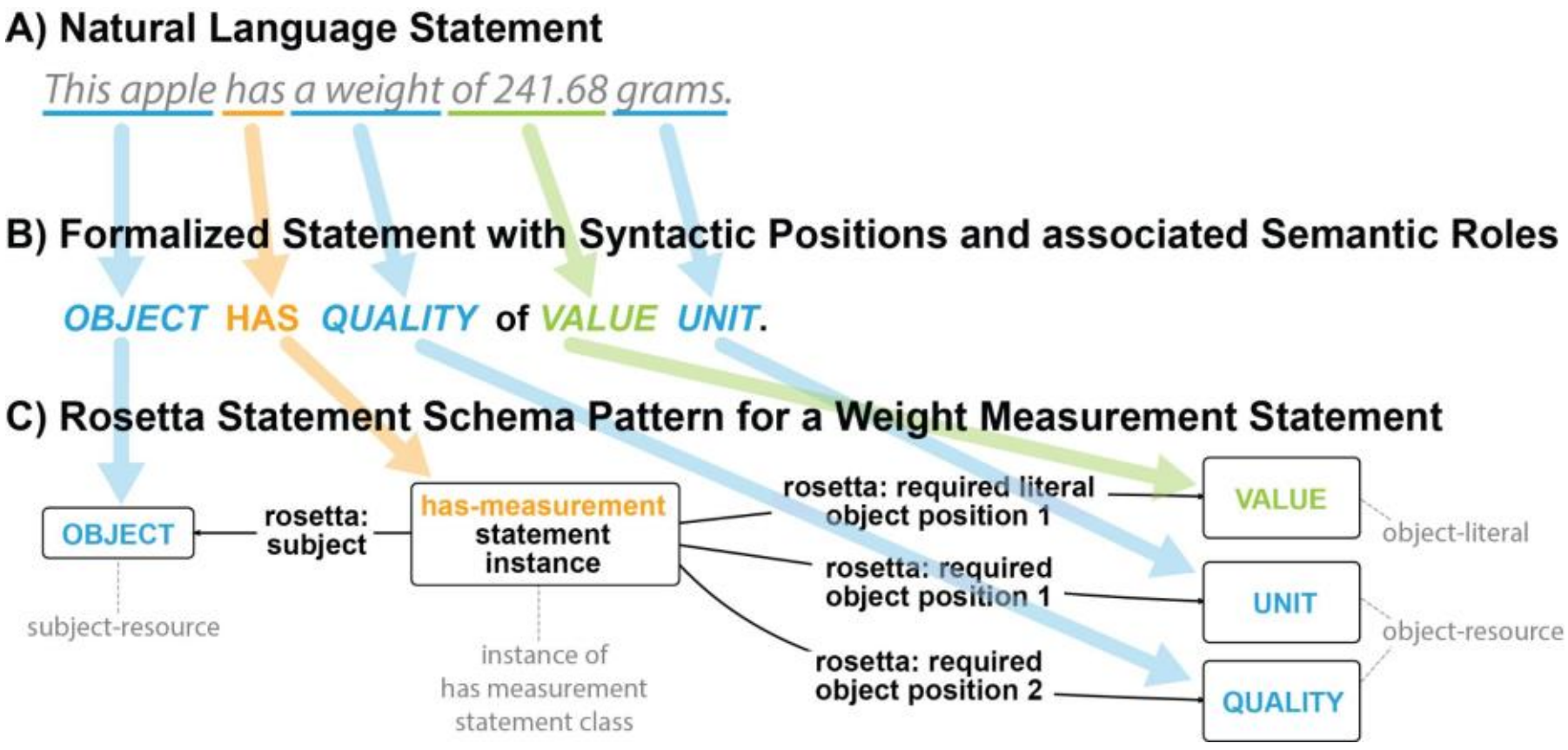

**Figure 4: From the structure of a natural language statement to the structure of a Rosetta Statement schema pattern. A)** A natural language statement (token model) with the predicate *has (measurement)*. **B)** The corresponding formalized statement (metamodel), with the syntactic positions and their associated semantic roles highlighted in color. **C)** The Rosetta Statement schema pattern (metamodel) for the has-measurement statement. Arrows indicate the alignment between positions and slots across the three models.

The schema would also require two additional arguments to be added: (i) a value of 241.68 with datatype xsd:float as the object-literal and (ii) a named-individual resource '*gram*' (wikidata:Q41803) as the object-resource. The schema links the statement instance resource to these object-arguments via a sequentially numbered property (Fig. 4). To further streamline this pattern, it would be possible to eliminate the distinction between object-arguments and object-adjuncts, and instead utilize the property '*object position #*' to link the statement instance resource to the respective objects.

A comparison of this schema with the measurement schemata from OBO and OBOE (cf. Fig. 3) reveals that, on the one hand, fewer triples are required to model the statement—i.e., three instead of five or six—and on the other hand, much fewer classes are required. The Rosetta Statement schema pattern is characterized by its simplicity, containing only input slots and devoid of superfluous positions such as '*scalar measurement datum*' and '*scalar value specification*' in the OBO schema or

[17] or any other ontology class that refers to apples.

'*observation*' and '*measurement*' in the OBOE schema. These additional positions and their associated resources hold no relevance for a human reader, who is solely interested in the information contained within the input slots, and are therefore not covered by the Rosetta Statement. The additional positions are also not relevant for translating the semantic content back into a natural language statement that is semantically meaningful to a domain expert.

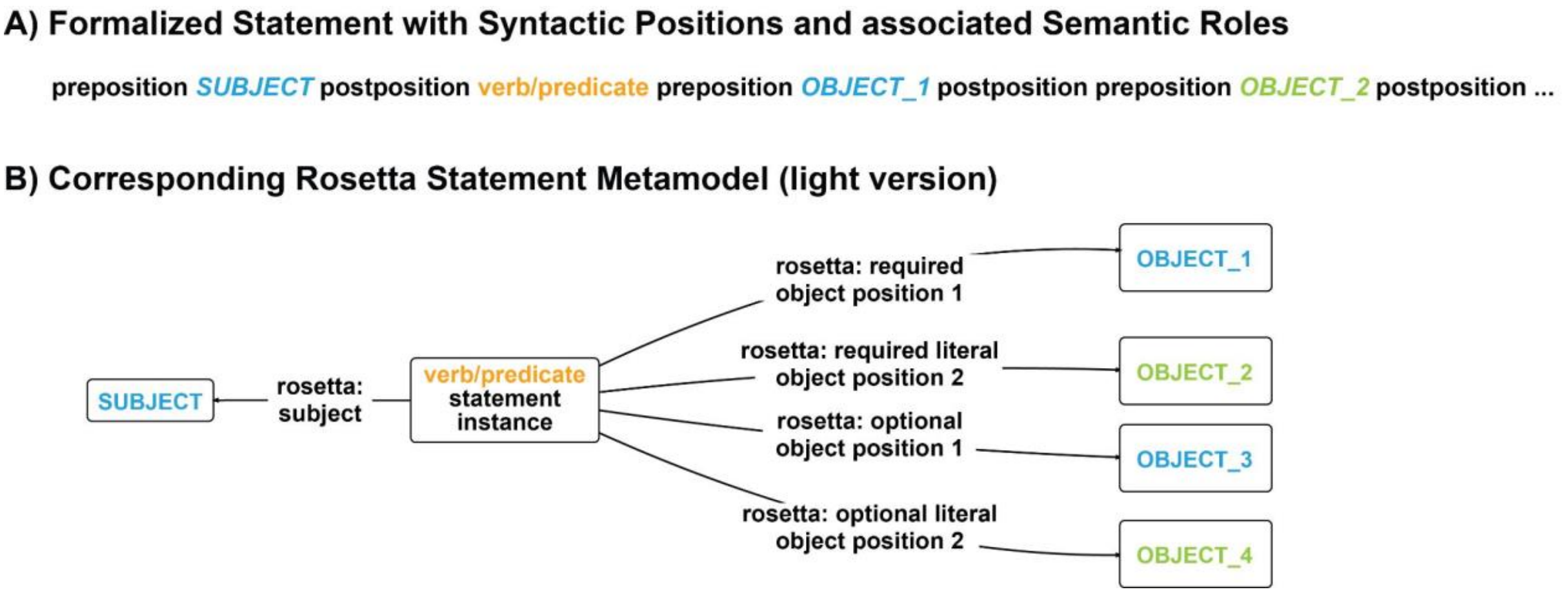

**Figure 5: From a formalized natural language statement to the corresponding light version of the Rosetta Statement metamodel. A)** A formalized statement with its syntactic positions and associated semantic roles highlighted in color. **B)** The light version of the Rosetta Statement metamodel from A). The statement instance resource indirectly indicates the verb or predicate of the statement, shown in orange. Object arguments ('*rosetta:required object position #*' for resources and '*rosetta:required literal object position #*' for literals) and adjuncts ('*rosetta:optional object position #*' for resources and '*rosetta:optional literal object position #*' for literals) can be either object-resources (in blue) or object-literals (in green).

The Rosetta Statement modelling approach can be applied to any simple English statement consisting of a single verb or predicate (see Figure 5), and it invariably generates statements that are represented in the graph by their own dedicated resource that instantiates a corresponding semantic data schema pattern that belongs to the corresponding Rosetta Statement class. Consequently, one can make statements about each Rosetta Statement without having to apply RDF reification (53) or RDF-star (54,55), which are feasible for referring to individual triples but not for larger subgraphs such as a measurement datum with a 95% confidence interval (see Fig. 1, middle), for which the latter two approaches are inefficient and complicated to query. Named Graphs emerge as another potential solution for such larger subgraphs (53), and one could organize all triples belonging to a Rosetta Statement into its own Named Graph using the GUPRI of the statement instance resource as the GUPRI of the Named Graph. However, employing Named Graphs is not required, as Rosetta Statements can always be referred to via their statement resource.

In essence, the Rosetta Statement approach to knowledge graph construction employs the notion of RDF reification, albeit applied to natural language statements rather than reifying a triple. This renders the modelling of n-ary statements straightforward, as well as statements concerning n-ary statements. Conversely, the mind-independent reality modelling approach often poses a substantial modelling challenge for n-ary statements. A similar predicament arises when attempting to formulate statements about statements via RDF reification. This challenge is further compounded when multiple triples must be reified to model a statement, as illustrated in Figure 1. The Rosetta Statement approach offers a practicable solution to this challenge.

As each instance of a Rosetta Statement class represents the statement as a whole, including its verb or predicate, one can use the statement resource to make statements about that Rosetta Statement, including statements (i) about the provenance of the statement, such as creator, creation date, author, curator, imported from, etc., (ii) about the GUPRI of the Rosetta Statement schema pattern that the statement instantiates (which can be specified as a SHACL shape), (iii) about the copyright license for the statement, (iv) about access/reading restrictions for specific user roles and rights for the statement, (v) about whether the statement can be edited and by whom, (vi) about a specification of the confidence level of the statement, which is of particular importance in scientific contexts (56,57), where a lack of confidence can lead to issues such as citation distortion (58), (vii) about a specification of the time interval for which the statement is valid, and (viii) about references as source evidence for the statement, to name a few possibilities.

Each argument in a given Rosetta Statement schema pattern can be aligned with a particular syntactic position, which is modelled in the schema as a slot. For each slot, the corresponding semantic role is specified as a **constraint specification**—either as an XML Schema datatype specification for an object-literal, which can be supplemented with a specific pattern or range constraint, or as an ontology class specification for a subject or an object-resource, which restricts the type of resources that can be located in a particular slot to that class or any of its subclasses. Corresponding Rosetta Statement schema patterns can be specified as **SHACL shapes**, for example. Statements modeled according to the same shape are **semantically interoperable** and **machine-interpretable statements**.

It is noteworthy that a given Rosetta Statement schema pattern can be extended to include additional object adjuncts, which can be added without causing any compatibility issues with previously created statements using older versions of the pattern. This is due to the fact that object adjuncts are considered optional and, as such, are not subject to the requirements of the reference schema.

## The full version of the Rosetta Statement metamodel, supporting versioning and the tracking of an editing history

As previously mentioned in the problem statement, some knowledge graphs, such as the ORKG, possess an open, domain agnostic scope and adhere to the DKGC approach. These graphs undergo rapid evolution, with their content being the product of collaborative or even crowdsourced editing. This editing process enables any user to modify any statement within the graph, including statements created by other users. For these specific knowledge graphs, it is imperative to possess the capability to track the editing history at the level of individual statements. This facilitates transparency and fosters trust. The incorporation of a citation mechanism for individual statements within the knowledge graph would serve to enhance its functionality as a valuable resource for scholarly communication. This addition would facilitate the organization and preservation of cited content, ensuring its accessibility and integrity over time. The integration of a versioning system within the knowledge graph would enable continuous evolution through user contributions, while maintaining the integrity of citations and references.

The statement versioning mechanism of the full version of the Rosetta Statement metamodel supports this, and it also supports tracking the editing history for each individual Rosetta Statement and each particular object-position[18]. However, this requires certain adaptations to the light version of the Rosetta Statement metamodel.

[18] The editing history does not cover any edits of individual resources, though, such as changing the label of a named-individual resource, which would affect every Rosetta Statement that has this resource in its subject or object position. The editing history of individual resources is not within the scope of the Rosetta Statement approach and requires its own solution.

In the full version of the Rosetta Statement metamodel (Fig. 6), subject-resources, object-resources, and object-literals are not directly linked to the statement instance, but indirectly through instances of a **subject-position class** and **object-position classes**. Whereas the subject-position class can be reused in any Rosetta Statement, independent of the statement type, the object-position classes are defined for each object argument and adjunct position of each Rosetta Statement class. Consequently, each particular Rosetta Statement of a given statement type has, in addition to an instance of the corresponding Rosetta Statement class, an instance of each object-position class, to which the actual object-resources and object-literals are linked, and an instance of the general subject-position class to which the subject-resource is linked. Consequently, one can refer to every subject and object of every Rosetta Statement individually by the GUPRI of its subject- and object-position resource.

The number of object-position classes that a given Rosetta Statement pattern distinguishes depends on the n-arity of the underlying statement type. The dependency of object-position classes on their corresponding statement type is documented within the respective Rosetta Statement class as a class axiom that points to the required and optional object-position classes.

This structure also supports having more than one subject-resource and object-resource or object-literal in a given position with the same semantic role, allowing to make statements such as '*Sarah and Anna* met *Bob and Christopher* on *4th of July 2021* in *New York City*' or '*Anna and Bob* travel by *train* from *Berlin* to *Paris* via *Osnabrück, Hengelo, Utrecht, and Rotterdam* on the *21st of April 2023*'. The order of the object and subject resources aligning with the same semantic role is specified via a '*rosetta:order*' property, followed by a sequentially increasing integer.

By introducing the notion of an **anchor statement resource** to which different statement version resources can be linked via a '*rosetta:has version*' property, the metamodel supports both the versioning of statements and tracking the editing history for each object position (Fig. 6). The anchor statement resource represents the statement independent of its version and must always be resolved by the knowledge graph application to the newest statement version available. Like any of its statement version resources, it instantiates the respective Rosetta Statement class (e.g., travelling statement class). It points to the Rosetta Statement pattern specification that it instantiates via the property '*rosetta:has data schema pattern*'. Via the property '*rosetta:has context*' (inverse relation: '*rosetta:has statement*'), the statement can be linked to other content in the knowledge graph, such as the scholarly publication from which the statement has been taken.

Various metadata can be associated with the anchor statement resource, indicating the creator, creation date, the author, the extraction method (e.g., if the statement has been extracted from text by machines), from where the statement has been imported (if applicable), and whether the statement should be modifiable to users of the knowledge graph or be unchangeable. If soft-delete of statements should be supported, so that a Rosetta Statement is still in the graph when a user "deletes" it, the statement is only marked as "deleted" at the level of the anchor statement resource via using the properties '*rosetta:deleted at*' and '*rosetta:deleted by*'. The backend of the knowledge graph application will process this information and may still provide the provenance metadata associated with the "deleted" statement, but not the statement itself. With this, knowledge graphs based on Rosetta Statements also fulfill principle A2 of the FAIR Principles, requiring metadata to be accessible, even when the data are no longer available (5).

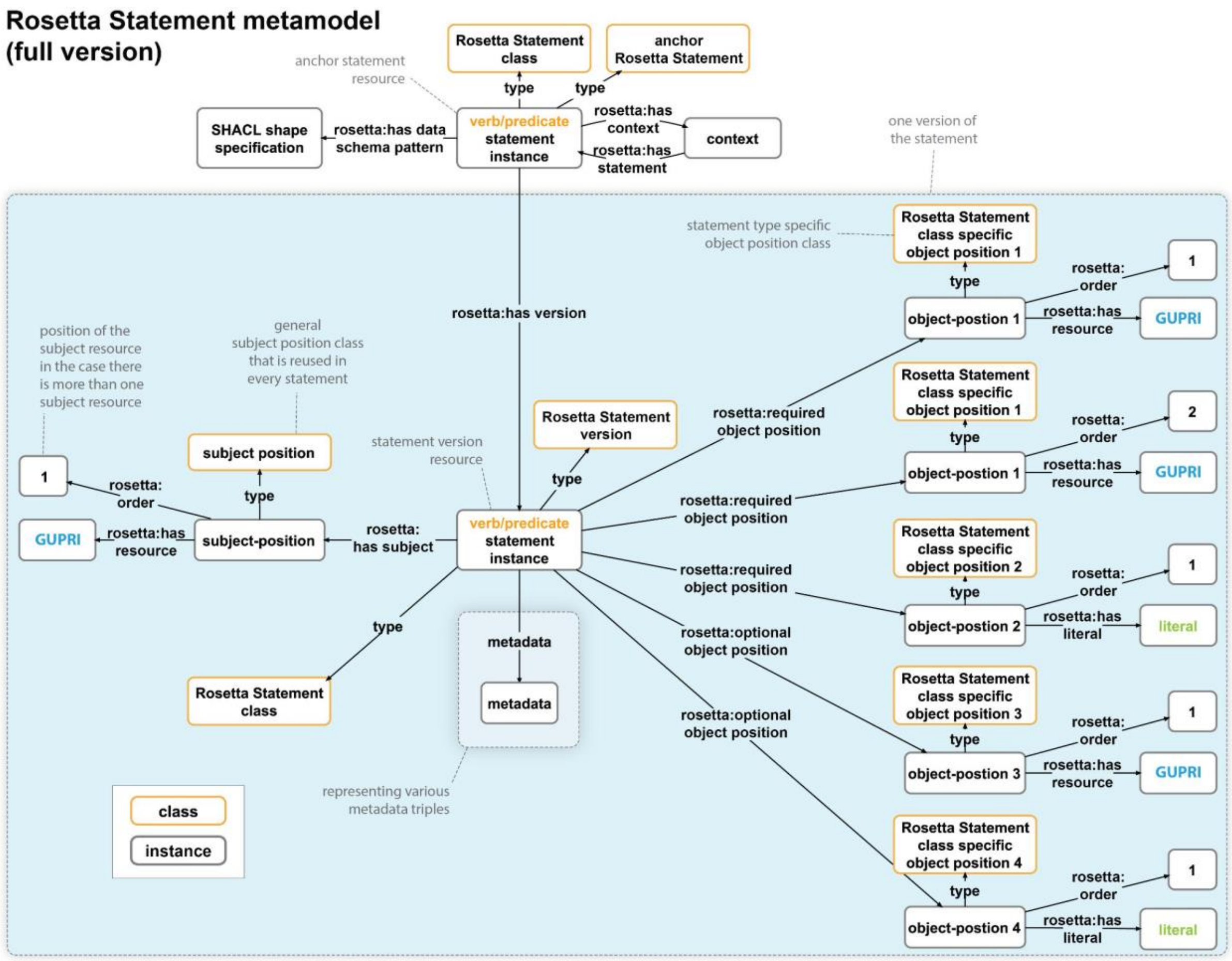

**Figure 6: Structure of the full version of the Rosetta Statement metamodel.** The Rosetta Statement pattern for the statement from Figure 5 A), according to the full version of the Rosetta Statement metamodel. Compared to the light version (Fig. 5B), it introduces the possibility of having several statement versions by linking each statement version resource to an anchor statement resource via a '*rosetta:has version*' property. The anchor resource specifies an optional context to which the statement belongs (e.g., a scholarly publication) and identifies through the '*rosetta:has data schema pattern*' property the Rosetta Statement semantic data schema pattern that it instantiates. Each version has a statement instance to which, indirectly, a number of objects and subjects are linked. Indirect, because each Rosetta Statement class has, depending on the arity of its statement, one or more accompanying object-position classes defined—one for each object argument and adjunct. For a given statement version, the corresponding object-position classes are instantiated and linked to the statement version instance, depending on whether they are arguments ('*rosetta:required object position*') or adjuncts ('*rosetta:optional object position*'). The actual object-resources (blue 'GUPRI') and object-literals (green 'literal') are linked to their respective object-position instance. The same applies to the subject-resource, with the only difference that a general subject position class is used for all Rosetta Statements, independent of their type. This structure supports linking more than one subject resource and more than one object resource or literal to a given subject and object position. Since various metadata can be linked to each statement version resource, including the information that it has been (soft) deleted, the full version of the Rosetta Statement metamodel also supports the versioning of statements and the tracking of the editing history for each object position of each statement in a knowledge graph. Whenever a position is updated, a new version is created in the graph. *Metadata associated with the anchor statement resource is not shown.*

Each **statement version resource** represents a complete Rosetta Statement, together with accompanying metadata for this version, indicating the creator, creation date, the author, a specification of the statement's certainty, and a version identifier. If single versions should be deletable, '*rosetta:deleted at*' and '*rosetta:deleted by*' metadata would be linked to the statement version resource as well. The combined metadata of each individual version of the history of a Rosetta Statement forms the contributor metadata for the latest version of the Statement, and the creation date of the latest version is its last update date.

Based on this information, gathered from all versions of a given Rosetta Statement, all information necessary for providing the **editing history** of that statement is available, even for the editing history of individual subject and object positions (i.e., statement slots). And since every subject and object of a given statement version has its own GUPRI, one can refer to them individually. The version identifier, which could be a DOI, allows citing this specific version, while the Rosetta Statement may continue to evolve in the knowledge graph due to users updating it. Whenever a user updates a statement, a new version of the statement is created and linked to the anchor statement resource, with each statement version resource having its own consecutive version number. The version with the highest version number is the latest and therefore current version of the statement.

If the distinction between object arguments and adjuncts and thus between required and optional objects is not desired, the metamodel can be simplified, using only the property '*rosetta:has object position*' to link a version statement resource to its object-position resources.

Moreover, if some basic rule-based reasoning should be supported, one can also specify that a particular **logical property** of the verb or predicate of the statement applies to a particular object-position by using a corresponding Boolean annotation property (e.g., '*rosetta:transitive*') with the respective object-position instance. This way, it would be possible, for example, to document that the transitivity of a has-part statement applies to the resource specified for the PART object-position.

The versioning and editing history, like it is defined for the full version of the Rosetta Statements metamodel, provides semantically structured information that can be used by humans and machines to monitor the "evolution" of a dynamic knowledge graph, identifying typical change-chains and hot topics, trends etc. The information can also be utilized for optimizing the UI and knowledge graph structure.

It is important to note that **the Rosetta Statement approach does not claim to model and represent a human-independent reality**, as other approaches to semantic modelling attempt to do, such as the Open Biological and Biomedical Ontology (OBO) Foundry, with the Basic Formal Ontology (BFO) (59) as its top-level ontology. Instead, it follows a **pragmatic approach with a focus on the efficient and reliable communication of information of all kinds** between humans and machines and across machines, including but not restricted to (meta)data statements. For the time being, the approach is limited to terms and statements as meaning-carrying units of information, but can be extended to larger units in the future (see discussion below).

And as an aside, due to the generic structure of the Rosetta Statement metamodel, its application for representing semantic content is not limited to knowledge graphs, but can also be applied to data structures for relational databases.

## Rosetta Statement display templates

Humans usually do not want to see the semantic content of a knowledge graph in the form of triples—they do not want to read them in any of the RDF serializations, nor do they want to visualize them as an RDF/OWL graph (see *Cognitive interoperability challenge*). In order to display semantic content that is modeled according to the Rosetta Statement approach in a human-actionable way, a frontend application needs a **display template** that is associated with every type of Rosetta Statement and that specifies how a rendering function can translate semantic content in the knowledge graph into a human-readable statement. Display templates organize information from a Rosetta Statement graph along with additional pre- and postpositions to be presented in the UI.

For example, if the measurement statement with 95% confidence interval from Figure 1 were stored according to a corresponding Rosetta Statement pattern, the pattern would specify four literal-object-positions (*MAIN_VALUE*, *UPPER_VALUE*, *LOWER_VALUE*, *INTERVAL_VALUE*), two resource-object-positions (*QUALITY*, *UNIT*), and one resource-subject-position (*OBJECT*). The textual display template could specify that this information should be displayed in the frontend as 'This *OBJECT* has a *QUALITY* of *MAIN_VALUE UNIT* (*INTERVAL_VALUE*% conf. interval: *LOWER_VALUE-UPPER_VALUE UNIT*)'. In the case of the weight measurement statement from Figure 1, this would read "*This apple* has a *weight* of *241.68 grams* (*95%* conf. interval: *241.31-242.05 grams*)". Another example is a travel statement where the corresponding Rosetta Statement pattern specifies one required resource-object-position (*DESTINATION_LOCATION*), two optional resource-object-positions (*DEPARTURE_LOCATION*, *TRANSPORTATION*), and one optional literal-object-position (*DATETIME*). Together with the subject-position (*PERSON*), the corresponding display template would display a travel statement in the frontend as '*PERSON* travels by *TRANSPORTATION* from *DEPARTURE_LOCATION* to *DESTINATION_LOCATION* on the *DATETIME*'. In other words, the subject-position and the various object-positions (i.e., the syntactic positions with their associated semantic roles) are mapped to corresponding variables within a string to form a human-readable statement (see Fig. 7, top). We call such textual display templates **dynamic labels**.

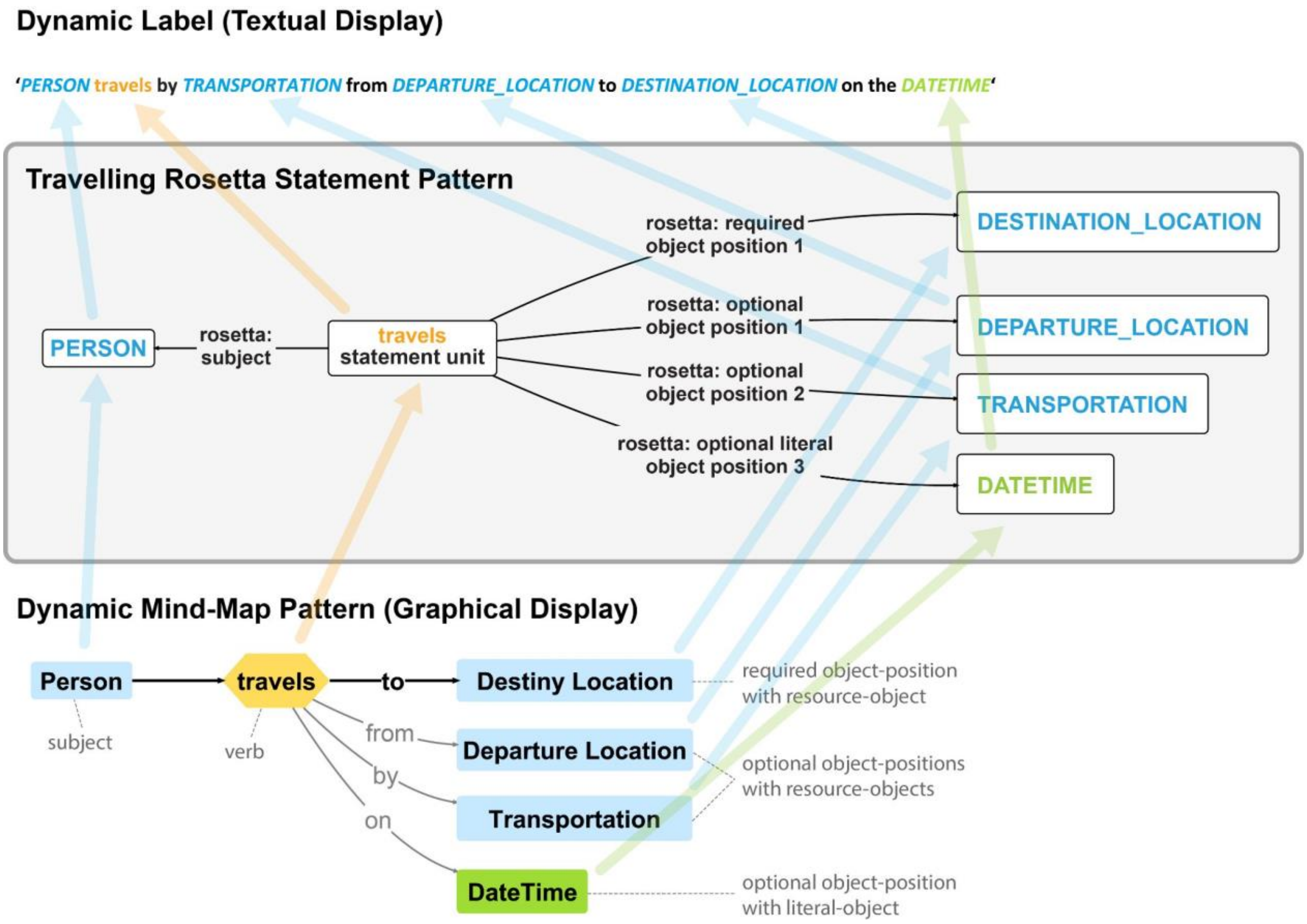

**Figure 7: Textual and graphical displays of a statement based on its Rosetta Statement pattern. Middle:** The Rosetta Statement pattern of a travels statement (here, applying the light version of the Rosetta Statement approach). **Top:** A textual display of the travels statement, i.e., a dynamic label that is associated with the Rosetta Statement pattern. Note how the subject-position and the object-position slots from the pattern align (arrows) with variable-positions (i.e., syntactic positions with associated semantic roles) in the dynamic label template. **Bottom:** A graphical display of the traveling statement, i.e., a dynamic mind-map pattern that is associated with the Rosetta Statement pattern. The alignment of the subject-position and the object-position slots from the Rosetta Statement pattern to nodes in the dynamic mind-map pattern is similar to the alignment for the dynamic label template.

In addition to textual display templates, it is also possible to specify graphical display templates for a mind-map-like representation of a statement using **dynamic mind-map patterns** (Fig. 7, bottom). Dynamic mind-map patterns use a label for the predicate underlying the corresponding statement type and, if there is more than one object-position, labels for relating the various objects to the predicate. As a result, the statement can be visualized as a mind-map like graph, where each subject and object is represented as a node with the label of the corresponding resource from the underlying Rosetta Statement graph. Such graphical representations of statements can also be combined through shared resources in subject- and object-positions, resulting in a mind-map of larger contexts and interrelationships that connect the dynamic mind-map patterns of multiple statements. Mind-map-like representations of complex interrelationships between different entities are often easier to understand than form-based textual representations, thus increasing the human-actionability of a knowledge graph—users do not want to read about family relationships but rather see the family tree.

Semantic content from the knowledge graph can be communicated to the presentation layer in the UI, with the templates filtering the complex data structure for the information relevant to a human user, using dynamic labels and dynamic mind-map patterns to present statements, **decoupling human-readable data display from machine-actionable data storage**.

A given Rosetta Statement pattern can have multiple dynamic labels and dynamic mind-map patterns associated with it. It can be beneficial to be able to choose between different display templates depending on the context in which the semantic content of a knowledge graph is accessed (PC vs. smartphone, expert user vs. layperson, etc.). The specifications of each display template should be associated with its corresponding Rosetta Statement class and pattern specification.

## Rosetta Statements and semantic interoperability: Specifying schema crosswalks

Different controlled vocabularies and ontologies may contain terms that have the same referent and sometimes even the same intensional meaning, in which case they would be strict synonyms. Unfortunately, if their GUPRIs differ, a machine will not be able to recognize them as synonyms, and statements using such terms will not be interoperable because the terms they use are not terminologically interoperable. In such cases, entity mappings between the GUPRIs of terms that share the same referent and ideally also the same intensional meaning can establish terminological interoperability (12,60).

Analog to entity mappings for establishing terminological interoperability, **schema crosswalks** can be specified between any given Rosetta Statement pattern and other semantic data schema patterns to establish **schema interoperability**. As type or metamodels that serve a specific purpose, data schemata are usually optimized towards a specific data use, with the uses depending on specific research questions. Consequently, no general optimal schema exists for any given type of data statement and the use of different data schemata across different projects is inevitable (12). By specifying different schema crosswalks for a given Rosetta Statement pattern, researchers can use a schema that is optimized for the set of operations and tools relevant to their particular project and research topic, while making the semantic content they document schematically interoperable with all other statements created with that Rosetta Statement pattern and all schemata for which schema crosswalks have been specified (Fig. 8).

In addition to specifying schema crosswalks between different semantic data schema patterns, crosswalks can also be specified between different formats such as RDF/OWL, GraphQL, Python or Java data classes, JSON, and CSV. Since all of these formats must provide data slots for a given statement type that map to their positions and their associated semantic roles, mapping to non-graph-based formats should be analogous to mapping to graph-based formats (e.g., Fig. 8, top right). This takes the observation into account that FAIRness is not sufficient as an indicator of high (meta)data quality—the use of (meta)data often depends on its **fitness-for-use**, i.e., data must be available in appropriate formats that comply with established standards and protocols that allow their direct use, e.g., when a specific analysis software requires data in a specific format and schema.

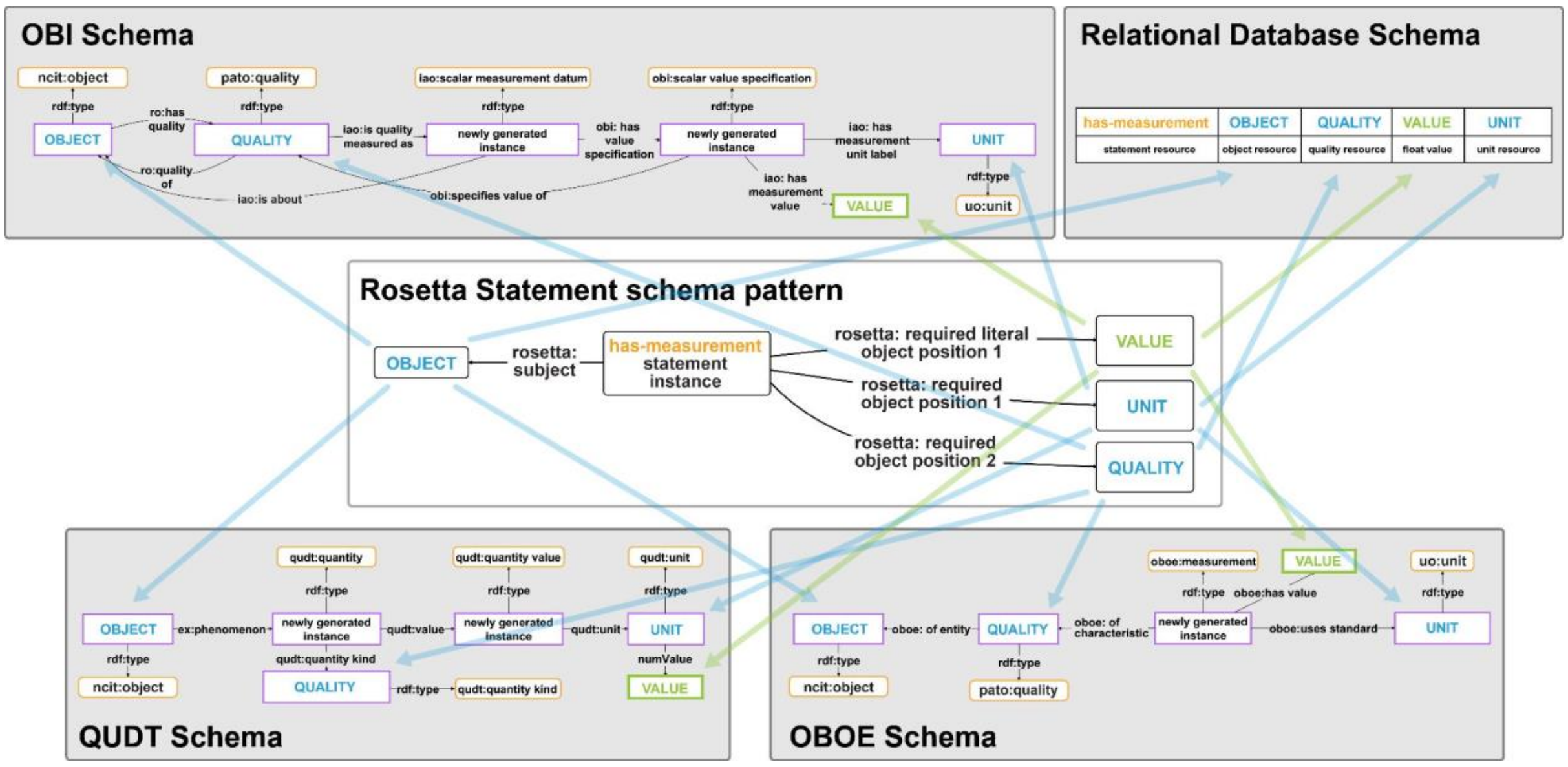

**Figure 8: Schema crosswalks from the light version Rosetta Statement schema pattern for a measurement statement to four other schemata**. **Middle:** The light version Rosetta Statement schema pattern for a measurement statement. **Top left:** The OBI schema. **Top right:** Relational database schema. **Bottom left:** The schema of the Quantity, Units, Dimensions, and Types Ontology (QUDT) (model transferred from (61)). **Bottom right:** The OBOE schema. Blue arrows indicate the alignment of object-resource slots and green arrows of object-literal slots. When operationalizing crosswalks, entity mappings are often required to satisfy the constraint specifications of the target schema. For example, if the OBI schema requires a class resource from the Units of Measurement Ontology (UO) for its UNIT slot, the Wikidata resource of the source must be translated into a corresponding UO resource, while for the QUDT schema it would have to be translated into a corresponding QUDT resource. Whereas the relational database model can cover the statement resource of a Rosetta Statement in a dedicated column of the table, the other three semantic data schema patterns would have to model the statement in a Named Graph, which would have the statement resource as its GUPRI.

The ability to specify schema crosswalks that convert, for example, measurement statements that comply with a corresponding Rosetta Statement pattern into data graphs that comply with the corresponding OBI schema also opens up the possibility for knowledge graph applications to establish workflows in which statements that meet certain criteria, such as having a certain confidence level or having a documented reference to a relevant source of evidence for the statement, are then converted into data graphs that comply with the OBI schema for measurements, thereby converting information from a schema that models statements into a schema that models a human-independent reality.

# Rosetta Statement use case: Open Research Knowledge Graph

We are implementing the Rosetta Statement approach in the ORKG. The ORKG is an open and domain agnostic knowledge graph infrastructure that supports the structured description of scholarly findings (i.e., scholarly semantic content) that were originally published in articles and books and thus expressed as narrative texts, tables, figures, and diagrams (i.e., textual representational artifacts). The ORKG follows the goal to provide these findings in a FAIR and machine-actionable format to support their reuse (33,34). To reach this goal, it applies the DKGC approach, with the challenges that this entails (see *Problem statement*), and complements this with semi-automated Large Language Model (LLM) and Natural Language Processing (NLP) approaches that guide users in the semantic parsing task of transforming and representing semantic content from scholarly articles in the knowledge graph (62). It also applies NLP tools for suggesting relevant pre-defined input templates to users (63). Its UI applies a semantic data browser, with which users can explore the graph along triple paths following RDF links, resulting in the cognitive interoperability challenge described above (see *Problem statement*).

Currently, users can add content to the ORKG by either (1) selecting a pre-defined semantic data schema pattern, i.e. an ORKG template, from which an input form is generated that the user completes (with a template editor that allows users to define new ORKG templates), or by (2) creating individual triples following the RDF syntax of *Subject-Predicate-Object* by specifying properties and linking to them a resource or literal as their objects. A 'contribution' resource, which is linked to the respective publication resource in the graph, serves as the subject for the initial triples. Additional triples can be created by linking to these initial triples, using resources from their *Object* position as the subject, and so forth. In doing so, the ORKG grants its users complete autonomy in the manner of how they model contributions, including defining their own properties and class terms but also reusing properties and class terms defined by other ORKG users or from existing ontologies. These two approaches can also be combined. Whereas the second approach imposes the task of semantic parsing and thus the parsing burden on the user (see *Semantic parsing burden challenge*), the first approach requires the pre-definition of ORKG templates for any type of semantic content users possibly want to add to the ORKG, which is practically impossible (see *Dynamic knowledge graph construction challenge*), and if users want to define their own ORKG templates, they are faced with the parsing burden again.

By implementing the Rosetta Statement approach, we provide users of the ORKG a third option for entering semantic content to the graph. We implemented the full version that supports versioning and tracking the editing history, and we distinguish between required and optional objects (i.e., object arguments and adjuncts) via the object-position resources. For now, we included only the dynamic label as a textual display of Rosetta Statements and not also the dynamic mind-map patterns for their graphical display.

When a user wants to add semantic content to the ORKG, instead of having to choose from one of the pre-defined ORKG templates, creating a new template, or adding a triple by choosing an existing property or creating a new one, the user now can add a statement by choosing a Rosetta Statement pattern from the list of existing patterns or define a new pattern.

The ORKG UI guides a user who wants to create a new Rosetta Statement pattern and its accompanying Rosetta Statement class through a form they have to complete. This procedure lowers the barrier and significantly reduces the semantic parsing burden for the users of ORKG (see *Semantic parsing burden challenge*).

The statement type editor asks for a label (Fig. 9A) and a description (Fig. 9B) for the new statement type, which provides the label and the definition of the new Rosetta Statement class. In the next input field, the user is asked to provide example natural language sentences, which will be documented in the Rosetta Statement class (Fig. 9C). The UI also provides an editing overview with a display of the dynamic label of the Rosetta Statement pattern at the top, based on the information available so far (Fig. 9D). At the beginning of this step, the dynamic label will only show a *Subject* and a *Verb* element. Below, the UI lists the subject position and the verb or predicate as expandable items (Fig. 9E), with the possibility to add object positions at the bottom. By expanding any of the items, further editing possibilities for each position become available. For the verb position, only the label can be specified, indicating how the verb is displayed in the dynamic label.

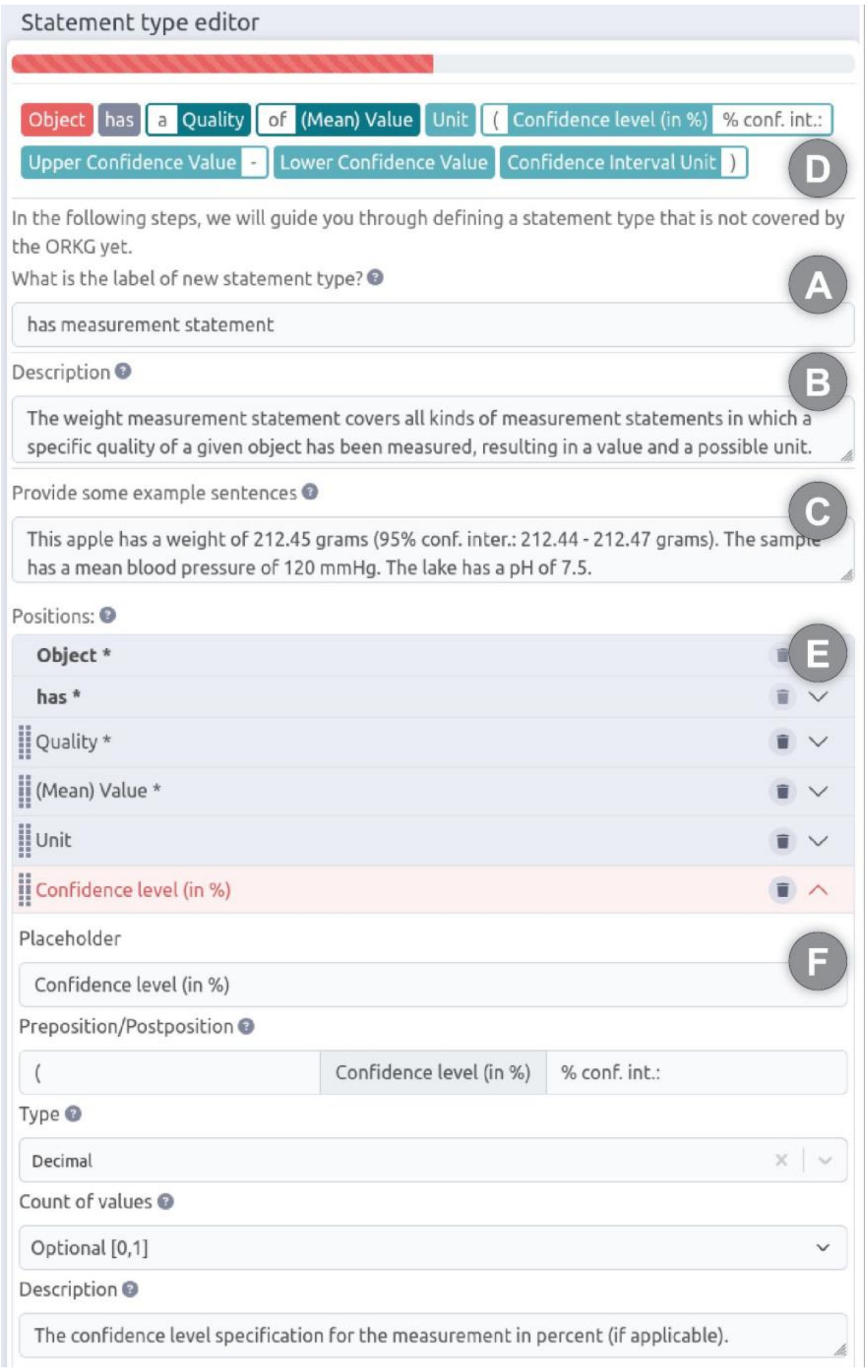

**Figure 9: Input form for specifying a new Rosetta Statement pattern and accompanying Rosetta Statement class**. Users provide a label (**A**) and a definition or description (**B**) for the corresponding Rosetta Statement class, together with some example sentences (**C**). An overview of the editing progress is provided (**D**), with initially only the subject position and the verb being specified (here, an advanced editing status is shown). More object positions can be added. Each position in the statement is represented in an item list (**E**), and items can be expanded to add more information to them (**F**), specifying their placeholder label, pre- and postposition labels, type restrictions (i.e., input constraints), specification of number of values allowed for the position, and additional text that describes what information is expected for the position (here shown for the Confidence Level position). How the information provided for each position influences the dynamic label textual display is directly shown (**D**), and users can change the order of the object positions by drag and drop of the items in the list (required object positions are displayed in darker and optional object positions in lighter color).

For all other positions, the following information can be entered (Fig. 9F):

- a placeholder text, which is displayed in the corresponding input field when a user wants to add a new statement of this type (cf. Fig. 10C) and which also appears in the overview of the editing process (Fig. 9D);
- a pre- and postposition text that specifies the text that the dynamic label will display directly before and after the subject or object position, in case the position is not empty (cf. Fig. 10A,B);

- the type of input that is allowed for the given position and thus its constraint specification (e.g., *resource*, *integer*, *decimal*, *URL*, *Boolean*, *date*, or *text*)—currently, the ORKG templating system does not support restricting the input of a position to a specific ontology class and its subclasses;
- the count of values allowed for each position, providing the possibility to allow multiple subjects or objects for a given subject- and object-position;
- and a description for the position that provides additional information that is shown to users when they enter a statement using this Rosetta Statement pattern and that can be used in the case that the placeholder text, which needs to be short, is not informative enough.

All information relating to the display label that is entered during this editing step is directly visualized, so that users can see how their editing progress influences the dynamic label and thus how Rosetta Statements of that type will be displayed in the ORKG UI (Fig. 9D). Users can also change the order of the object positions during this editing step via drag and drop of position headers, and the dynamic label will update accordingly. The Rosetta Statement pattern along with the corresponding Rosetta Statement class can be saved by clicking the button '*Create and insert statement type*'.

Whenever a user wants to add a statement of that newly defined statement type, the UI creates a corresponding input form based on the corresponding Rosetta Statement pattern (Fig. 10C). Users can still reuse class terms from existing ontologies or define their own class terms, but they do not have to define their own properties anymore, since the Rosetta Statement metamodel uses a fixed set of general properties and models all other properties as Rosetta Statement classes. In the view mode, instead of having to explore a complex representation of the semantic content represented in the measurement statement using the semantic data browser (which would require 15 clicks to gather the relevant information for a measurement with confidence interval; see *Cognitive interoperability challenge*), the Rosetta Statement is displayed as a natural language statement by a single click, using the dynamic label associated with it (Fig. 10A). The display label also adapts to whether some positions are empty and does not display corresponding information (cf. Fig. 10 A with B).

The use of the Rosetta Statement semantic parsing paradigm results in Rosetta Statement patterns that are easily understood by domain experts because the underlying general data structure reflects the structure of English natural language statements. Furthermore, specifying Rosetta Statement patterns for new types of statements is not as demanding when following the Rosetta Statement modelling paradigm. Not only developers, but also domain experts (and anyone else) who want to use knowledge graph applications and who are not experts in semantics and knowledge modelling in particular, or computer science in general, will be able to specify new Rosetta Statement patterns. Moreover, contrary to the other approaches of adding content to the ORKG, each Rosetta Statement is always based on a corresponding Rosetta Statement schema pattern associated with a Rosetta Statement class. With the ORKG editor for new Rosetta Statement types, the need to develop semantic data graph patterns to establish FAIR (meta)data within the RDF framework will no longer be such a barrier (see *Semantic parsing burden challenge*), and we expect that the development of supporting tools will be more straightforward due to the Rosetta Statement metamodel being shared across all Rosetta Statement types.

As each Rosetta Statement is represented in the graph with its own resource and as an instance of a particular statement class, it is straightforward to make statements about these statements. This includes, in addition to the above-mentioned possibility to link all kinds of metadata to the statement resource, also the possibility to indicate the **modifiability** of each statement as a Boolean property. It also allows specifying the **degree of certainty** of a given statement, which represents important information that may contribute to preventing citation distortion (57,58). Both are supported by default in the ORKG.

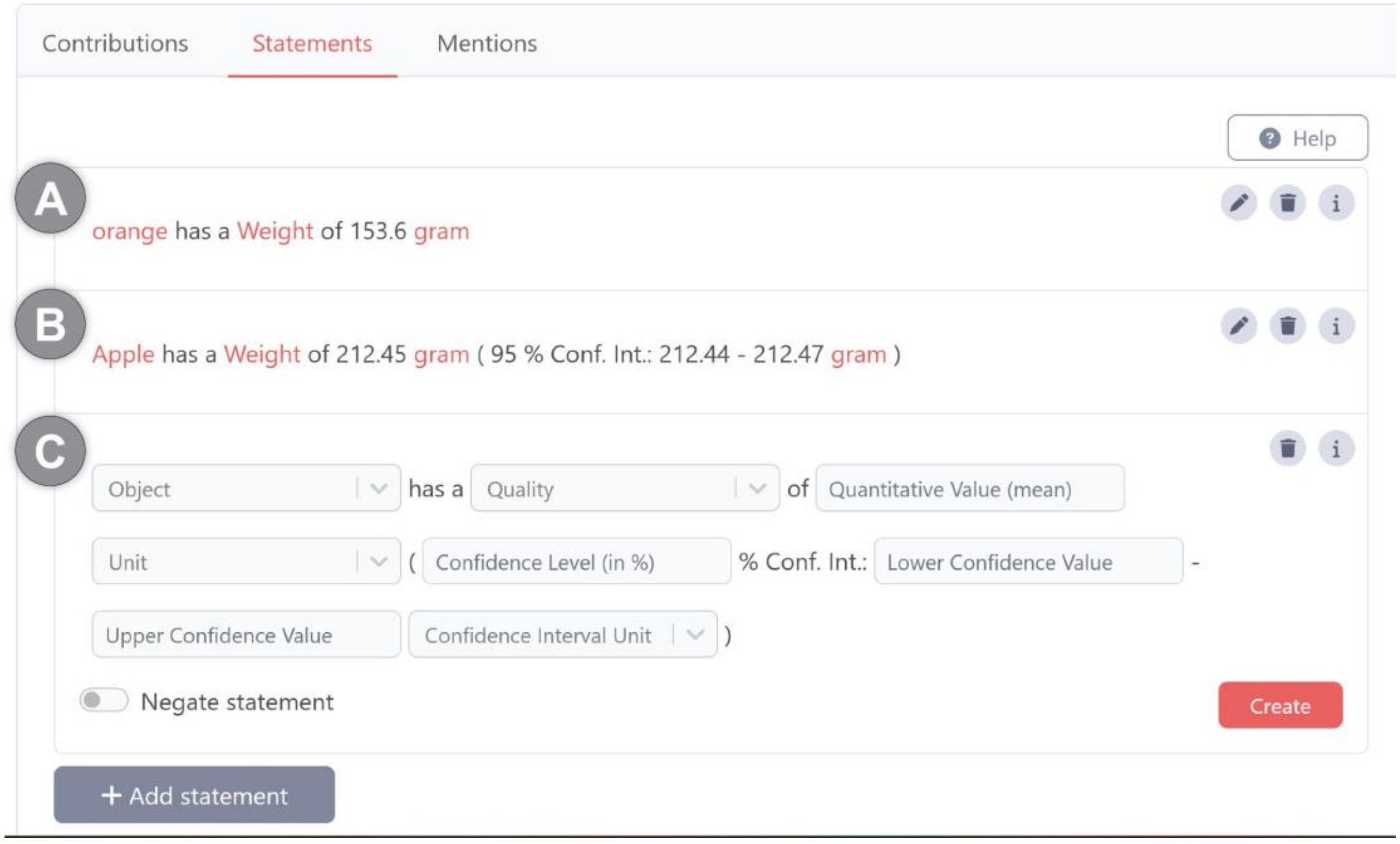

**Figure 10: Display of Rosetta Statements in the ORKG UI in the view and the edit mode. A:** The representation of a weight measurement statement of a particular orange using the *measurement statement* Rosetta Statement pattern, without specifying a confidence level. Empty object-positions and their associated pre- and postposition texts are not displayed in the view mode. **B:** The representation of a weight measurement statement of a particular apple, with all positions of the pattern being specified. **C:** The input form for the *measurement statement* Rosetta Statement in the ORKG. The subject-position ('Object') and each object-position defined in the corresponding Rosetta Statement pattern aligns with an input field in the ORKG UI when creating or editing the corresponding Rosetta Statement instance.

With Rosetta Statements, users can now also relate semantic content across different scholarly publications in the ORKG. They can, for instance, relate an observation stated in one paper to a hypothesis stated in another paper, specifying that the observation contradicts the hypothesis. Such cross-document referencing tasks form a significant part of reading and writing activities in scholarly research (64), and associating such information across different papers is challenging without the aid of digital tools (65).

Every Rosetta Statement is instantiating a corresponding Rosetta Statement class, which is defined in reference to its main verb/predicate. In addition to this verb/predicate-based classification, we introduced the possibility to classify any given Rosetta Statement as an instance of the '*rosetta:negation*' class, indicating that the corresponding statement is negated. This can be used to express typical **negations**, but also **absence statements**, which are often needed when describing specific objects, situations, or events. In OWL, following the Open World Assumption, modelling negations, including absences such as '*This hand has no thumb*', requires the specification of appropriate class axioms and blank nodes. By classifying statement resources as instances of a class '*rosetta:negation*', we follow the suggestion made in the context of semantic units (49,66) to model any statement as being negated by classifying it as a negation. Modelling negations in this way is considerably simpler and easier to comprehend by domain experts than modelling them as TBox expressions, and would thus increase their cognitive interoperability (see (11,66) for more details).

Modelling natural language statements instead of a mind-independent reality thus offers several advantages in the context of open, user-driven, and domain agnostic knowledge graphs such as the ORKG. We expect the Rosetta Statement approach to simplify the semantic parsing task, lowering the barrier for adding semantic content to the ORKG. With its shared metamodel, the approach also provides a universal and generally applicable graph data schema structure against which functions can be programmed without having to consider the various peculiarities of domain-specific semantic data schema patterns, which would be the case when modelling a mind-independent reality. Furthermore, the structural proximity of Rosetta Statements to natural language statements facilitates the involvement of LLMs in developing services and functions to support users in any task that involves interacting with Rosetta Statement graphs (see LLM-based support for creating Rosetta Statements and summarizing them). And although graphs based on the Rosetta Statement approach are not capable of reasoning, reasoning capability can still be achieved for parts of the ORKG graph by defining schema crosswalks between selected Rosetta Statement patterns and semantic data schema patterns that support reasoning.

## Rosetta Statement nanopublications as FAIR Digital Objects

As part of a strategy to accomplish FAIR (meta)data, it has been suggested to organize (meta)data into FAIR Digital Objects. A FAIR Digital Object is a digital object that is identifiable and resolvable by a GUPRI, such as a DOI, a handle, or an Internationalized Resource Identifier (IRI) (18,67).

Nanopublications can serve as FAIR Digital Objects. A nanopublication represents the smallest unit of publishable information extracted from literature and enriched with provenance and attribution information, documented in an RDF graph utilizing Named Graphs and Semantic Web technology (68–70). Nanopublications model specific assertions, such as scientific claims, using a machine-readable format and semantics, and each nanopublication is accessible and citable through a unique identifier, facilitating the discovery, exploration, and re-use of scholarly assertions (71). A nanopublication consists of four individual named graphs:

- The head graph describes the structure of the nanopublication itself, declaring the type of the nanopublication resource and connecting it to the other three named graphs.
- The assertion graph describes the main content of the nanopublication.
- The provenance graph describes metadata about the assertion itself, e.g., which scientific method was used to generate the assertion.
- The publication info graph describes additional metadata about the nanopublication, e.g., who created the nanopublication and when.

In the ORKG, Rosetta Statements can be published as nanopublications, with each Rosetta Statement version being represented as its own nanopublication. The assertion graph is based on the graph structure of the statement version, but without any metadata attached. Instead, the metadata of the Rosetta Statement is split across the provenance and publication info subgraphs, depending on the type of metadata. As a result, the implementation of the Rosetta Statement approach in the ORKG enables its users to publish individual statement FAIR Digital Objects in the form of nanopublications.

# Future work

We have several ideas for future improvements to the ORKG on the basis of the Rosetta Statement approach. In the following, we briefly discuss some of them.

### Templating Rosetta Statement forms

In many cases, users want to describe a specific type of entity in a standardized way, using the same set of statement types, and they want the respective semantic content to be organized in such a way that the corresponding Rosetta Statements are displayed in a specific order and grouped according to specific topics. For example, when information is standardized and consists of a specific collection of statements, some of which are required and others are optional according to the standard, such as in product data sheets in industry or disease report sheets in healthcare.

To support cases like these, we need an editor for specifying Rosetta Statement forms, which are templates of ordered collections of Rosetta Statement patterns. Each Rosetta Statement form template specifies the types of Rosetta Statements it covers, their cardinality (i.e., whether they are required and how many statements of that type are allowed), and their position within the form. Statements can be organized into groups, each with their own header phrase that will be displayed in the UI of the ORKG. The forms must also include links between slots across different Rosetta Statements so that for instance the object resource of statement *A* is automatically set to be the subject resource of statement *B*, resulting in connected statements that form a connected graph.

When a particular form is meant to be used to describe a specific type of entity, the respective ontology class should be referenced in the form template as well. This allows the ORKG UI to suggest users the form whenever they add an instance of that class to the graph.

### Rosetta Statements and semantic units

Rosetta Statement forms result in collections of semantically related Rosetta Statements. Just like a Rosetta Statement is represented in the graph with its own GUPRI that instantiates a specific Rosetta Statement class, each such collection of statements can also be represented by its own GUPRI, instantiating a corresponding Rosetta Statement collection class. The semantic content modelled by such a collection would be represented in the graph by its own resource, with the associated form template providing the corresponding semantic data schema pattern. Different such collection classes can be distinguished and organized in a taxonomy of Rosetta Statement collection class types. Since every collection in the graph would possess its own GUPRI that instantiates the corresponding class, the overall ORKG graph would be organized into various semantically meaningful subgraphs, enhancing the structure and navigability of the graph and facilitating context-dependent graph exploration. This would implement the concept of semantic units to the ORKG and would allow utilizing all the applications of semantic units discussed so far (11,19,49,66).

Furthermore, the semantic unit framework introduced three new types of resources: some-instance, most-instances, and every-instance resources (for a discussion, see (66)). These new resource types complement named-individual, class, and property resources. By adopting these new resource types and employing them in Rosetta Statements, we can formally represent and differentiate between assertional, contingent, prototypical, and universal Rosetta Statements, all of which can be represented as ABox expressions in the knowledge graph. The conventional OWL modelling only enables the formal representation and distinction of assertional and universal statements, but not of contingent and prototypical statements. Moreover, it does not allow the representation of universal statements within the knowledge graph (66)[19]. Semantic units, however, support their formal representation and distinction. The ORKG would benefit from supporting all four statement categories, as they play an essential role in scientific communication.

By implementing semantic units in the ORKG, many more classification criteria can be applied, leading to a sophisticated classification of different types of statements based on their meaning, their epistemic value and function, their referents, and their contexts, including time-indexed and geo-indexed statements, conditional if-then statements, granularity trees, disagreement, logical arguments (deduction, induction, abduction), cardinality restrictions, directive statements, and questions (49,66).

## A Rosetta Statement search and exploration interface

Since every Rosetta Statement is associated with a specific Rosetta Statement class and its accompanying semantic graph pattern, we want to support ORKG users in searching and exploring all content in the ORKG that is based on Rosetta Statements. We want to provide the following two workflows:

1. **Search by specific term**: Users should be able to search for a specific term and get a list of Rosetta Statements sorted by statement types that include this term in their subject or object positions.
2. **Search by first selecting a specific Rosetta Statement type and then using facets**: Users should be able to search for a specific type of Rosetta Statement. When they have selected the Rosetta Statement type they are interested in, they can use facets provided by the interface to explore and narrow down the search result. Each subject and object position that the semantic graph pattern defines for the selected Rosetta Statement type is represented by its own facet.

This enables users of the ORKG to query the content in the ORKG that is based on Rosetta Statements without having to write queries with a graph query language (see *Graph query challenge*).

## LLM-based support for creating Rosetta Statements and their summarized displays

As all Rosetta Statement patterns are instantiations of the same underlying metamodel, all semantic content in the ORKG that is modelled as Rosetta Statements is structured in the same way. This facilitates the development of tools and services that interact with Rosetta Statements, as these tools and services can be developed against that common structure. It also facilitates the development of tools that utilize LLMs approaches.

Recent studies have demonstrated the considerable potential of LLMs to enhance the general accessibility of scientific knowledge (25–27,72–75). While still requiring a human-in-the-loop for final evaluation, LLMs also have proven to provide substantial (semi-)automated support in the creation of knowledge graphs and ontologies based on input texts (28–31). These approaches, however, typically require the pre-definition of relevant semantic graph patterns to achieve semantic interoperability and thus do not circumvent the semantic parsing burden. Moreover, in the domain of science and research knowledge graphs, where precision, subtlety, and data reliability are paramount, our experience suggests that the current capabilities of LLMs to support the typically applied approaches for knowledge graph creation are limited.

[19] In OWL, universal statements can only be documented in an ontology as class axioms of ontology classes in the form of TBox expressions that involve blank nodes. Consequently, they are not part of the universe of discourse of a knowledge graph.

The structural alignment between the Rosetta Statement metamodel and the structure of simple English natural language sentences, however, suggests that LLMs are well suited to process them. Consequently, we anticipate the productive utilization of LLMs in supporting tools for the extraction of Rosetta Statements from input texts and for the specification of new Rosetta Statement types and their accompanying Rosetta Statement patterns. We envision the following LLM-supported workflow:

1. Along with the input text, the following information must be provided:
    a. A list of Rosetta Statement classes that correspond with the types of information to be extracted.
    b. The formalized natural language statement metamodel associated with each Rosetta Statement class of the list (step 1a). The formalized statements can be derived from the dynamic labels. For example, '*PERSON* travels from *DEPARTURE_LOCATION* to *DESTINATION_LOCATION* by *TRANSPORTATION* on *DATE*' for a travels-statement class. The different syntactic positions and their corresponding thematic labels must be listed as variables for each formalized statement. This would be *PERSON*, *DEPARTURE_LOCATION*, *DESTINATION_LOCATION*, *TRANSPORTATION*, and *DATE* for the travels-statement example.
    c. The SHACL shape specification for each Rosetta Statement class from the list.
    d. The ontologies that are referenced in the constraint specifications of the various slots of the SHACL shapes (step 1c).
2. A prompt is specified that asks the LLM
    a. to identify information in the input text that corresponds with the formalized natural language metamodels (step 1b),
    b. to translate passages from the input text that contain this information into correspondingly structured statements using the formalized statements as templates, and
    c. to return the text passages alongside with these structured statements.
3. A subsequent prompt takes the outcome of step 2 and asks for identifying terms in the ontologies identified in step 1d that match the terms used in the structured statements, thereby meeting the constraint specifications for each statement position. Following our example, for the *DEPARTURE_LOCATION* position, it would, for instance, only allow locations from the GeoNames Ontology.
4. The next prompt translates the structured statements from step 2, with the ontology terms identified in step 3, into an RDF graph based on the corresponding SHACL shape from step 1c.
5. The RDF graphs from step 4 are finally validated against their corresponding SHACL shapes using a SHACL validator. This should identify any deviations the LLM made (hallucinations, etc.) from the semantic data schema pattern specified by the shape. It should also identify any made up ontology resources, as they would violate the slot constraints of the shape.
6. In the last step, the extracted Rosetta Statements are displayed in the UI alongside with the corresponding text passages from the input text for final approval by the user.

It must be acknowledged that this is an experimental approach, and it will be necessary to evaluate the extent to which this will provide support for ORKG users. For the specification of new Rosetta Statement types, it could also be useful to utilize existing resources such as The Berkeley

FrameNet (76,77) dataset that comprises specifications for over 1,200 semantic frames, each of which includes the specification of possible subject and object positions (i.e., frame elements) and possible verbs and predicates (i.e., lexical units), together with a body of 200,000 manually annotated sentences. We expect that with such kind of data, coupled with latest LLM approaches, we can develop tools that will substantially support a workflow for (semi-)automatically specifying new Rosetta Statement types, with associated semantic data schema patterns and dynamic labels. Such LLM-based tools would substantially enhance the overall usability of the ORKG (78).

Regarding the presentation of semantic content from Rosetta Statements in the ORKG as a result of a query, we want to experiment with the capability of LLMs to summarize collections of statements (i.e., Rosetta Statements) into a cohesive and well readable text, but also plan to consider other approaches (79).

# Discussion

Applying the Rosetta Statement semantic parsing approach organizes a knowledge graph into a set of statements, each of which represents a minimum information unit that is semantically meaningful to a human reader. This set of statements mathematically partitions the graph, with each triple in the graph belonging to exactly one Rosetta Statement, and therewith adds another layer to the knowledge graph, above the layer of triples. One can criticize the need of having to add such a layer, but it also allows organizing all content of a knowledge graph into a set of nanopublications (68–70), with each nanopublication being a FAIR Digital Object that documents a particular Rosetta Statement. Moreover, semantically meaningful collections of Rosetta Statements can be organized as semantic units in the form of RO-Crates and thus FAIR Digital Objects of a coarser granularity, adding more layers of coarser representational granularity to the knowledge graph (see also discussion in the context of semantic units (19,49,66). As a consequence, however, the Rosetta Statement approach requires the specification of a Rosetta Statement class with associated schema pattern for each type of statement to be added to a knowledge graph, which could be seen as a point of criticism. This limitation results from the fact that in the Rosetta Statement approach, the set of statement classes and associated schema patterns defines the possible **proposition-space** of the knowledge graph. While the criticism is valid, we want to respond that for a truly FAIR knowledge graph, every (meta)data statement must be FAIR. For a statement to be FAIR, it must be interoperable, and we explained above why this requires the specification of a schema pattern for each statement type. To guarantee schema interoperability and thus to be truly FAIR, each statement must also reference the identifier of the schema pattern against which it was modeled, and ideally also which statement type it instantiates (12). The necessity of ensuring the transparency and replicability of data modelling in a FAIR knowledge graph is not exclusive to the Rosetta Statement approach; it is applicable to all FAIR knowledge graphs. The Rosetta Statement framework offers a straightforward approach to the quality of semantic parsing that is required for establishing FAIR knowledge graphs (see also *Semantic parsing burden challenge*).

Regardless, since it is ultimately essential that the communication of information between machines and humans is efficient and reliable, and since humans communicate textual information using natural language statements, **semantically meaningful statements** should be the **building blocks** of any knowledge graph. With the Rosetta Statements approach, we suggest a framework for creating such knowledge graph building blocks that meet the cognitive requirements of humans and facilitate their ways of communication. With the concept of semantic units, we can combine Rosetta Stone building blocks to form larger, semantically meaningful collections of statements.

The distinction between the Rosetta Statement approach for creating a knowledge graph, such as implemented in the ORKG, with the Rosetta Statement type editor, and other knowledge graph frameworks lies in the capacity of the former to facilitate the creation of schemata for new statement types without requiring any expertise in Semantics, RDF, OWL, or graph query languages. Consequently, the barrier to developing a knowledge graph is being significantly reduced. Open knowledge graphs with a cross-domain scope that use the DKGC approach for entering semantic content to the graph benefit in particular from the Rosetta Statement approach, as applying the static approach to dynamic scenarios or domains usually falls short when a new type of statement needs to be added to the graph, which would shift the semantic parsing burden onto the user.

It would be unreasonable to expect the Rosetta Statement approach to resolve all semantic interoperability issues that arise from a DKGC approach. There will be cases of two or more Rosetta Statement patterns having been created for the same type of statement. For instance, the statement "*Peter travels by bus from Berlin to Paris*" could also be expressed as "*Peter took a bus to get from Berlin to Paris*". We think that a combination of providing sufficient synonymous predicates and verbs to define the underlying Rosetta Statement class and the description of a general statement theme or frame could potentially reduce the number of such semantically overlapping Rosetta Statement classes, and data exist that can be utilized for this task (see, e.g., The Berkeley FrameNet). Regardless, since every pattern complies with the same underlying metamodel, curating the corresponding semantic content in a knowledge graph becomes considerably easier than when such a common underlying data structure is missing. Also, the Rosetta Statement approach reduces the number of properties that have to be defined to model semantic content to the ones used in the Rosetta Statement metamodel and its associated metadata. Combined with the fact that all the Rosetta Statement schema patterns comply with the Rosetta Statement metamodel significantly reduces interoperability issues that otherwise most certainly would occur, and also supports the development of curation workflows to manage them. One could for instance identify Rosetta Statements that have been created using different patterns but still have a significant overlap in specific subject and object positions. They would be candidates for curators to take a look at and possibly merge to a single pattern.

Another criticism is that the Rosetta Statement semantic parsing paradigm does not relate to a logical framework, so you cannot apply reasoning to statements created with it. We agree that the ability to apply reasoning is a valuable asset. However, we consider it more crucial to prioritize the findability and overall FAIRness and cognitive interoperability of the semantic content of a knowledge graph. This is the reason why we developed the Rosetta Statement semantic parsing approach, which models natural statements and does not attempt to model a mind-independent reality. If reasoning is required, Rosetta Statements can always be converted into a structure that enables reasoning using corresponding schema crosswalks (see Fig. 8).

With the Rosetta Statement approach, we can now follow a three-step procedure for semantic parsing (see Fig. 12). In the first step, all semantic content to be represented in a FAIR knowledge graph can be modelled following the Rosetta Statement approach, using Wikidata as the underlying controlled vocabulary. This comes with the benefit of significantly lowering the entry barrier to adding content to a knowledge graph. In a next step, by providing entity-mappings between the Wikidata terms used in the knowledge graph and corresponding terms from OWL ontologies and by specifying slot-constraints in reference to these ontology terms for the Rosetta Statement patterns used in the knowledge graph, the graph can be transferred to support semantic search, therewith significantly increasing its search-capabilities (i.e., increasing the findability of (meta)data). Finally, in the third step, statement types for which reasoning support is desired can be identified. Only for this part of the knowledge graph, semantic data schema patterns that model a mind-independent reality must be defined, which requires the expertise of ontology engineers. By specifying schema crosswalks between Rosetta Statement schema patterns and reasoning-supporting semantic data schema patterns, all semantic content that should be reasoned about can be exported into reasoning-supporting graph structures.

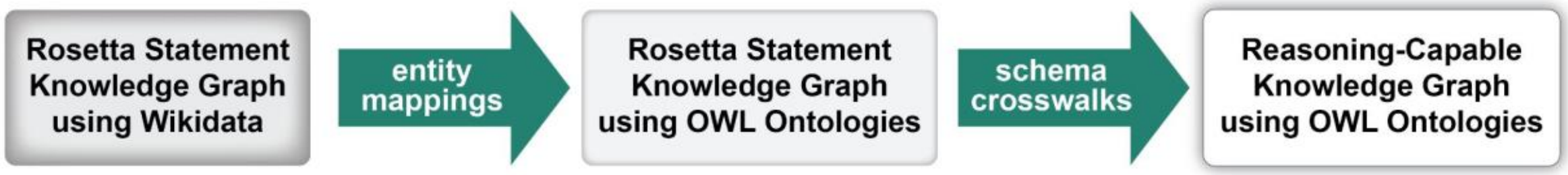

**Figure 12: Step-wise procedure for knowledge graph construction to lower the barrier from the semantic parsing burden.** In the first step, the Rosetta Statement Approach to knowledge graph construction is applied, using Wikidata as the underlying controlled vocabulary, resulting in a FAIR knowledge graph with limited findability. In a next step, by defining entity mappings between the Wikidata terms used in the knowledge graph and corresponding resources from established OWL ontologies, the FAIR knowledge graph can be transferred, now supporting semantic search. By developing reasoning-supporting semantic data schema patterns for each Rosetta Statement schema pattern used in the knowledge graph and by defining schema crosswalks between them, the knowledge graph can be transferred into a reasoning-capable FAIR knowledge graph in the final step.

If the Rosetta Statement schema patterns were made available in a schema repository, alongside with any associated schema crosswalks and operational functions, they could be reused by anyone developing their own knowledge graph applications. They could be used as knowledge graph building blocks by domain experts to create their own knowledge graphs for their various research projects.

By applying the Rosetta Statement approach in the ORKG, we also lower the barrier of reusing its semantic content within third-party applications, following the notion of a *System of Systems*, i.e., a collaborative and interactive information ecosystem that is continually evolving with its building blocks being defined functionally rather than concretely and for which new applications can be created at any time (80). Rosetta Statements with their semantic data schema patterns represent the building blocks of that system that can be created, shared, and discovered, that can collaboratively evolve and that guarantees a high degree of interoperability.

# Conclusion

To enable meaningful insights and fact-based decision-making, we need to harness machine support to integrate disparate datasets that are hidden in project-specific data silos. This requires the datasets to be interoperable across the projects. We have argued that only truly **FAIR and machine-actionable (meta)data** can support this objective, with ontologies, knowledge graphs, and semantic graph patterns being promising candidate concepts and technologies to achieve this. Without a way to make project-specific datasets interoperable, we will have a hard time coming up with practicable solutions for the major global challenges of biodiversity loss, zoonotic diseases, and climate change, all of which require a truly interdisciplinary approach.

In the *problem statement*, we identify four main challenges that we think represent major obstacles for achieving true FAIRness and machine-actionability of (meta)data across different projects. With the Rosetta Statement approach to knowledge graph construction, we introduce a metamodel for modelling semantic content in a knowledge graph that reflects the structure of English natural language statements. Based on this approach and in combination with the concept of semantic units (19,49,66), we think we can contribute to solutions for these challenges.

When modelling a mind-independent reality using the RDF syntax of *Subject-Predicate-Object*, we end up with representations that may be machine-actionable in the sense that they support reasoning, but they are often not readily comprehensible to the people who produce or want to use the data—the data structures often **lack human-actionability because they lack cognitive interoperability** (see *Cognitive interoperability challenge*; Fig. 1). With Rosetta Statements, (meta)data are represented in a way that reflects the **structure of English natural language statements** and, thus, should be significantly easier to comprehend and reuse. In addition, with the **dynamic labels**, Rosetta Statements provide a way to display their semantic content as natural language sentences in UIs.

When domain experts want to search for a particular (type of) semantic content within a knowledge graph, they are often asked to engage with a SPARQL or Cypher endpoint, through which they can write graph queries using the respective graph query language. The need to write such queries is a barrier to interacting with a knowledge graph, thus limiting the practical findability of its content (see *Graph query challenge*). Rosetta Statements facilitate a user-friendly approach to searching and exploring the contents in a knowledge graph, utilizing **facets for each subject and object position** of a given type of Rosetta Statement. This approach does not require users to have knowledge in semantics and graph query languages.

Another challenge that we identified is that in order to guarantee the **semantic interoperability** of (meta)data in a knowledge graph, domain experts must closely collaborate with ontology engineers to develop the ontology terms and semantic data schema patterns that are required for defining the respective **proposition-space** of the knowledge graph. This is very time-consuming and not practically achievable, especially for smaller projects. We called it the **semantic parsing burden** (see *Semantic parsing burden challenge*). With the ORKG use case of a knowledge graph that employed the Rosetta Statement approach, we can show that domain experts without any experience in semantics and data modelling can create their own **Rosetta Statement schema patterns,** guided by the ORKG Rosetta Statement type editor. This provides a first level of structured representations of (meta)data in a knowledge graph, using **Wikidata** as an underlying general terminology. This approach can be further improved by defining for each Rosetta Statement type a corresponding SHACL shape with specified constraints for each subject and object position, using XML Schema datatype specifications for object-literals, supplemented with specific patterns or range constraints, and ontology class specifications for subject- and object-resources, restricting the type of resources that can be used in a particular slot to instances of that class or any of its subclasses. This would allow small teams that lack expertise in semantics to avoid the semantic parsing burden but still build FAIR knowledge graphs for their projects, either using Wikidata as the underlying terminology, or, if they have a basic understanding of semantics and ontologies in their domain, using domain ontologies and specifying corresponding SHACL shapes for their Rosetta Statements.

Due to their structural similarity to natural language statements, we expect **LLM approaches** to provide a promising framework for developing **supporting tools** for adding content to knowledge graphs using Rosetta Statement patterns, to create new Rosetta Statement patterns, and to display and also summarize content from larger collections of Rosetta Statements.

Finally, we are convinced that open knowledge graphs with a **domain agnostic scope**, such as the ORKG, that therefore follow a **community-driven DKGC approach**, will significantly benefit from employing the Rosetta Statement approach (see *Dynamic knowledge graph construction challenge*). A **Rosetta Statement schema pattern editor** like the one employed in the ORKG allows any user to specify a new Rosetta Statement schema pattern on the basis of the general Rosetta Statement metamodel. No knowledge and experience in semantic parsing is required, and thus the users are liberated from the semantic parsing burden, and the knowledge graph service providers do not have to pre-define all semantic data schema patterns that could possibly be required by a user—which is, as we have argued, not feasible anyway in a community-driven DKGC approach. And Wikidata can be used as an underlying general terminology.

We argue that the Rosetta Statement approach provides a framework in which the entry barrier for adding and finding semantic content in a knowledge graph is substantially lowered. Furthermore, it allows for a **three-level strategy towards developing digital twins**, with the first level being semantic content in the form of Rosetta Statements using **Wikidata** as terminology. The resulting knowledge graph would **meet basic criteria for findability and FAIRness**. For the second level, Wikidata terms are replaced by terms from an **OWL ontology** (e.g., by defining appropriate entity mappings). This would enable **semantic search** and thus increases the findability and search functionality of the resulting knowledge graph. In the third level, parts of the graph or the graph as a whole can be transformed into **reasoning-supporting** semantic data schema patterns by specifying respective schemata and corresponding schema crosswalks (see Fig. 12).

With the Rosetta Statement approach, we attempt to put cognitive interoperability as an essential criterion at the center of our design. We attempt to provide solutions that are usable for all kinds of **knowledge graph users**, including **knowledge graph builders** (e.g., domain experts with project data, developers of databases), **data analysts** (e.g., data scientists attempting to integrate various datasets, machine-learning experts searching for high-quality training data), and **consumers** (e.g., domain experts searching for data relevant to their research question) (20). We are aware of the fact that this approach ends up being a different way of thinking and building knowledge graphs than what is the norm, but we believe it is essential for achieving true FAIRness and machine-actionability of (meta)data, to think together the societal, cognitive, and interdisciplinary barriers and requirements for using knowledge graphs, and find appropriate pragmatic solutions. Intermediate steps must be taken towards the overall goal of research to create models of the world that are real enough to be useful.

In knowledge graph design, we tend to focus on the problem to be solved—providing a model of the world that can be reasoned over—when working on solutions, but we often do so outside the context in which the problem is embedded, which includes the domain experts, who we expect to use the knowledge graph. Sometimes, this results in us providing solutions that may solve the focus problem, but, at the same time, creating new problems down the line with cognitive interoperability. When creating solutions, we have to take into account the whole picture, including the practical problems that domain experts face when using knowledge graphs or that developers have when setting up new knowledge graphs. With the Rosetta Statement approach, we want to contribute to solutions that support cognitive interoperability of knowledge graphs and their semantic content so that they can have an impact on research.

# Acknowledgements

We thank Peter Grobe for discussing some of the presented ideas. We are solely responsible for all the arguments and statements in this paper. Lars Vogt received funding by the ERC H2020 Project

'ScienceGraph' (819536) and by two projects funded by the German Federal Ministry of Education and Research (BMBF), i.e., DIGIT RUBBER (Grant no. 13XP5126B), and InSuKa (Grant no. 13XP5196F).